\documentclass[reprint,superscriptaddress,amsmath,amssymb,aps,prd,notitlepage,longbibliography,floatfix,nofootinbib,onecolumn]{revtex4-1}

\usepackage{tensor}     
\usepackage{mdframed}
\usepackage{graphicx}   
\usepackage[
colorlinks=true,        
citecolor=blue,         
linkcolor=blue,         
urlcolor=blue           
]{hyperref}             
\usepackage{bm}         
\usepackage{xcolor}     
\usepackage{array}      
\usepackage{lipsum}
\usepackage{color}      
\usepackage[utf8]{inputenc} 
\usepackage[section]{placeins} 
\usepackage{multirow}

\usepackage{physics}
\usepackage{subfigure}
\newcommand{\nc}{\newcommand*} 

\nc{\al}{\alpha}
\nc{\s}{\sigma}
\nc{\kp}{\kappa}
\nc{\dt}{\delta}
\nc{\Dt}{\Delta}
\nc{\Ld}{\Lambda}
\nc{\p}{\partial}
\nc{\Gm}{\Gamma}
\nc{\om}{\omega}
\nc{\Om}{\Omega}
\nc{\rd}{\mathrm{d}}
\nc{\Od}{\mathcal{O}} 
\def\({\left(}
\def\){\right)}
\def\[{\left[}
\def\]{\right]}
\def\e{\begin{equation}}
\def\q{\end{equation}}
\def\m{\begin{eqnarray}}
\def\n{\end{eqnarray}}
\nc{\Eq}[1]{Eq.~\eqref{#1}}     
\nc{\Eqs}[1]{Eqs.~\eqref{#1}}     
\nc{\Fig}[1]{Fig.~\ref{#1}}     
\nc{\Table}[1]{Table~\ref{#1}}  
\nc{\Sec}[1]{Sec.~\ref{#1}}     
\nc{\Msun}{M_\odot}             
\nc{\fpbh}{f_{\mathrm{PBH}}}    
\nc{\fpbhn}{f_{\mathrm{pbh0}}}    
\nc{\mR}{\mathcal{R}} 
\nc{\seq}{\sigma_{\mathrm{eq}}}
\nc{\ogw}{\Omega_{\mathrm{GW}}}
\nc{\gpcyr}{\mathrm{Gpc}^{-3}\,\mathrm{yr}^{-1}}
\nc{\lvc}{LIGO-Virgo} 
\nc{\SNR}{\mathrm{SNR}} 
\nc{\mmin}{{m_{\mathrm{min}}}}
\nc{\mmax}{{m_{\mathrm{max}}}}
\nc{\Mmin}{{M_{\mathrm{min}}}}
\nc{\fmin}{{f_{\mathrm{min}}}}
\nc{\VT}{\mathrm{VT}}
\nc{\rhoGW}{\rho_{\mathrm{GW}}}
\nc{\vth}{\vec{\theta}}
\nc{\vd}{\vec{d}}
\nc{\vla}{\vec{\lambda}}
\nc{\Nobs}{N_{\mathrm{obs}}}
\nc{\av}[1]{\langle #1 \rangle} 
\nc{\km}{\mathrm{km}}
\nc{\Mpc}{\mathrm{Mpc}}
\nc{\Tobs}{T_{\mathrm{obs}}}
\nc{\Ntemp}{N_{\mathrm{temp}}}
\nc{\fyr}{f_{\mathrm{yr}}}
\nc{\addref}{[\textcolor{red}{add ref}] } 
\nc{\eg}{\textit{e.g.~}}
\nc{\app}{\approx}
\nc{\hf}{\frac{1}{2}}
\nc{\discuss}{\textcolor{red}{Add discussion here!}}
\nc{\red}[1]{\textcolor{red}{#1}}

\begin{document}
	
\title{Gravitational Waves Induced by Scalar Perturbations with a Broken Power-law Peak}
	
\author{Chong-Zhi Li}
\affiliation{CAS Key Laboratory of Theoretical Physics, 
		Institute of Theoretical Physics, Chinese Academy of Sciences,
		Beijing 100190, China}
\affiliation{School of Physical Sciences, 
		University of Chinese Academy of Sciences, 
		No. 19A Yuquan Road, Beijing 100049, China}
	
\author{Chen Yuan}
\email{Corresponding author: chenyuan@tecnico.ulisboa.pt}
\affiliation{CENTRA, Departamento de Física, Instituto Superior Técnico – IST, Universidade de Lisboa – UL, Avenida Rovisco Pais 1, 1049–001 Lisboa, Portugal}
\author{Qing-Guo Huang}
\email{Corresponding author: huangqg@itp.ac.cn}
\affiliation{CAS Key Laboratory of Theoretical Physics, 
		Institute of Theoretical Physics, Chinese Academy of Sciences,
		Beijing 100190, China}
\affiliation{School of Physical Sciences, 
		University of Chinese Academy of Sciences, 
		No. 19A Yuquan Road, Beijing 100049, China}
\affiliation{School of Fundamental Physics and Mathematical Sciences
		Hangzhou Institute for Advanced Study, UCAS, Hangzhou 310024, China}

\date{\today}
	
\begin{abstract}
We give an analytical approximation for the energy spectrum of the scalar-induced gravitational waves (SIGWs) generated by a broken power-law power spectrum, and find that both the asymptotic power-law tails and the intermediate peak contribute distinct features to the SIGW spectrum. Moreover, the broken power-law power spectrum has abundant near-peak features and our results can be used as a near-peak approximation that covers a wide range of models. Our analytical approximation is useful in the rapid generation of the SIGW energy spectrum, which is beneficial for gravitational wave data analysis.
\end{abstract}
	
\maketitle

\section{introduction}
Gravitational waves (GWs) originating from cosmological processes offer an important approach to investigate the early universe, as they propagate almost freely in the cosmic background. Among various GWs, scalar-induced gravitational waves (SIGWs) are well-known as a secondary effect sourced by linear order scalar perturbations  \cite{Tomita:1967wkp, Matarrese:1992rp, Matarrese:1993zf, Matarrese:1997ay, Noh:2004bc, Carbone:2004iv, Nakamura:2004rm}. Additionally, SIGWs are closely connected with the formation of primordial black holes (PBHs) in the very early Universe \cite{Saito:2008jc,Saito:2009jt,Bugaev:2009zh,Bugaev:2010bb}.

According to the detection of the Cosmic Microwave Background (CMB) anisotropies \cite{Planck:2018vyg,Planck:2018jri}, the amplitude of the scalar power spectrum is constrained to be approximately $A_s \approx 2.1\times 10^{-9}$ across scales from megaparsecs (Mpc) to gigaparsecs (Gpc), exhibiting a nearly scale-invariant feature, which is too small to generate detectable SIGWs. 
However, the information at scales smaller than Mpc lies beyond the reach of CMB detection, leaving the possibility of significant enhancement of primordial scalar perturbations.
In this context, SIGWs can provide constraints on primordial scalar power spectrum, inflationary models, and the expansion history of the universe 
\cite{Ananda:2006af,Baumann:2007zm,Assadullahi:2009jc,Inomata:2018epa,Lu:2019sti,Yuan:2019udt,Kapadia:2020pnr,Ragavendra:2021qdu,Braglia:2020taf,Witkowski:2021raz,Balaji:2022dbi,Alabidi:2013lya,Inomata:2019zqy,Inomata:2019ivs,Domenech:2020kqm,Dalianis:2020gup,Fu:2019vqc,Kawasaki:2019hvt,Espinosa:2018eve,Kohri:2018awv,Domenech:2019quo,Unal:2018yaa,Ota:2020vfn,Yuan:2019wwo,Yuan:2020iwf,Zhang:2020uek,Atal:2021jyo,Adshead:2021hnm,Yuan:2023ofl,Ferrante:2022mui,Orlofsky:2016vbd,Byrnes:2018txb,Fu:2019vqc,Kawasaki:2019hvt,Meng:2022low}.

Moreover, if a large scalar perturbation exceeds a critical value, the overdense region  will collapse to form a PBH once the corresponding wavelength reenters the Hubble horizon.
In this case, the abundance of PBHs is closely related to the amplitude of the scalar power spectrum, and hence will be tightly constrained by the observations on SIGWs, leading to the significance of studying both PBHs and SIGWs as complementary topics \cite{Inomata:2016rbd,Nakama:2016gzw,Garcia-Bellido:2017aan,Sasaki:2018dmp,Clesse:2018ogk,Inomata:2020lmk,Papanikolaou:2020qtd,Domenech:2020ssp,Franciolini:2021nvv,Cang:2022jyc,Gehrman:2022imk,Papanikolaou:2022chm,Qiu:2022klm,Escriva:2022duf,Meng:2022low,Chang:2022nzu,Gehrman:2023esa,Ferrante:2023bgz,Gu:2023mmd,Yuan:2024yyo,Huang:2024wse,Papanikolaou:2024kjb,Papanikolaou:2024fzf}. For reviews on SIGWs, see \cite{Yuan:2021qgz,Domenech:2021ztg}.

Since the first detection of GW by LIGO, numerous GW events have been detected by the LIGO-Virgo-KAGRA collaborations \cite{LIGOScientific:2016aoc,LIGOScientific:2016dsl,LIGOScientific:2018mvr,LIGOScientific:2020ibl,LIGOScientific:2021usb,KAGRA:2021vkt}. It is possible that some of the events have primordial origins \cite{Sasaki:2016jop,Chen:2018czv,Raidal:2018bbj,DeLuca:2020qqa,Hall:2020daa,Bhagwat:2020bzh,Hutsi:2020sol,Wong:2020yig,DeLuca:2021wjr,Bavera:2021wmw,Franciolini:2021tla,Chen:2021nxo,Chen:2024dxh,Yuan:2024yyo,Huang:2024wse}, implying that the corresponding primordial perturbations would generate SIGWs in the frequency band of Pulsar Timing Arrays (PTAs) (see e.g., \cite{Yuan:2024yyo}).
Recently, hints of the stochastic GW background have been reported by the PTA collaborations \cite{NANOGrav:2023hde,NANOGrav:2023gor,NANOGrav:2023hvm,Reardon:2023gzh,Zic:2023gta,EPTA:2023sfo,EPTA:2023fyk,EPTA:2023xxk,Xu:2023wog}.
Alongside various astrophysical and cosmological explanations (see, for example, \cite{Bi:2023tib,Wu:2023hsa}), a potential interpretation is that this signal originates from SIGWs \cite{Franciolini:2023pbf,Inomata:2023zup,Liu:2023ymk,Yi:2023mbm,You:2023rmn,Jin:2023wri,Balaji:2023ehk,Basilakos:2023xof,Liu:2023pau,Basilakos:2023jvp}. 
Furthermore, SIGWs are expected to be detected by future space-based GW detectors such as LISA \cite{LISA:2017pwj}, Taiji \cite{Ruan:2018tsw}, TianQin \cite{TianQin:2015yph} and DECIGO \cite{Kawamura:2020pcg}. These observations will place constraints on the abundance of PBH potentially addressing the remaining mass window where PBHs could constitute a significant fraction of dark matter \cite{Bartolo:2018evs, Bartolo:2018rku, Carr:2020gox}.

Based on the semi-analytical formula derived in \cite{Espinosa:2018eve,Kohri:2018awv,Domenech:2019quo}, the SIGW energy spectrum has been studied for specific cases and some results will be relevant to this work.
During the radiation-dominated (RD) era, the infrared (IR) behavior is derived in \cite{Yuan:2019wwo}, assuming a general enhanced scalar power spectrum with both IR and ultraviolet (UV) cutoffs.
In this case, the GW spectrum scales as $k^3\ln^2 k$ if $k$ is much smaller than the width of remaining interval, while it shows a transition to $k^2 \ln^2 k$ if the width is extremely small.
As a special case, SIGWs arise from a log-normal scalar power spectrum were studied in \cite{Pi:2020otn} for the widths $\Delta\ll1$ and $\Delta\gtrsim1$ respectively.
In this paper, we analytically calculate the SIGW spectrum with the scalar power spectrum possessing a broken power-law feature with a smooth transition, which can typically be generated by single-field inflation models \cite{Ballesteros:2017fsr,Byrnes:2018txb,Carrilho:2019oqg,Atal:2021jyo}. This power spectrum also contains abundant features such as near-peak asymmetry and asymptotic power-law tails, which would lead to characteristic structures in the SIGW spectrum.

This paper is organized as follows. We briefly review the formula of SIGW spectrum during RD in Sec.~\ref{II}, and consider the broken power-law power spectrum in Sec.~\ref{III}. 
In Sec.~\ref{III.A}, we give the analytical result of SIGW spectrum generated by broken power-law power spectrum, and compare with the numerical result.
In Sec.~\ref{III.B} and Sec.~\ref{III.C}, we give the IR and UV behavior of our result.
Finally, we summarize our findings in Sec. \ref{IV}. To streamline the calculations, a quick approach can be taken by starting from the final result, Eq.~(\ref{ogw}), and referring back to the intermediate results as needed.

\section{SIGW during RD era}\label{II}
In this section, we will briefly review the SIGWs generated during the RD era. Consider only scalar and tensor perturbations, the perturbed Friedmann-Robertson-Walker (FRW) metric in Newtonian gauge can be written as
\e
\mathrm{d}s^2=a^2(\eta) \left[-(1+2\phi)\mathrm{d}\eta^2+
\left((1-2\phi)\delta_{ij}+\frac{h_{ij}}{2}
\right)\mathrm{d}x^i\mathrm{d}x^j\right],
\q
where $\phi$ is the linear scalar perturbation and $h_{ij}$ refers to the second-order tensor perturbation. For SIGWs in different gauges, see \cite{Tomikawa:2019tvi,DeLuca:2019ufz,Inomata:2019yww,Yuan:2019fwv,Giovannini:2020qta,Lu:2020diy}. 
During the RD era, the equations of motion for $\phi$ and $h_{ij}$ in Fourier space can be written as
\m
\phi_{\boldsymbol{k}} ''(\eta) +4\mathcal{H}\phi_{\boldsymbol{k}} '(\eta) +\frac{1}{3}k^2\phi_{\boldsymbol{k}}(\eta) &=& 0 ,\label{II.1}\\
h_{\boldsymbol{k},\lambda}''(\eta)+ 2\mathcal{H}h_{\boldsymbol{k},\lambda}'(\eta)+ k^2h_{\boldsymbol{k},\lambda}(\eta) &=& \mathcal{S}_{\lambda}(\boldsymbol{k},\eta).
\label{II.2}\n
The prime denotes the derivative with respect to the conformal time $\eta$. $\lambda$ denotes the two polarization modes of GWs, and $\mathcal{S}_\lambda$ is the source term up to $\phi_{\boldsymbol{k}}^2$ order. Here $\phi_{\boldsymbol{k}}(\eta)$ and $\mathcal{S}_\lambda(\boldsymbol{k},\eta)$ take the form:
\begin{gather}
\phi(\boldsymbol{k},\eta) \equiv \phi_{\boldsymbol{k}}
T_\phi (k\eta) = 
\phi_{\boldsymbol{k}} \frac{9}{(k\eta)^2}\left(
\frac{\sin{(k\eta/\sqrt{3})}}{k\eta/\sqrt{3}}- \cos{(k\eta/\sqrt{3})} \right),\\
\mathcal{S}_{\lambda}(\boldsymbol{k},\eta)=
- 4 \int\frac{\mathrm{d}^3 p}{(2\pi)^{3/2}}(e_\lambda^{i j} p_i p_j) \phi_{\boldsymbol{p}} \phi_{\boldsymbol{k-p}} F(p/k,|\boldsymbol{k}-\boldsymbol{p}|/k,k\eta) 
\label{II.3},\\
F(u,v,x) = 3 T_\phi(ux)T_\phi(vx) + ux T_\phi '(ux) T_\phi(vx) + vx T_\phi(ux) T_\phi '(vx)+u v x^2 T_\phi '(ux) T_\phi '(vx),    
\end{gather}
where $\phi_{\boldsymbol{k}}$ is the initial condition given by inflation models and $e_\lambda^{i j}$ is the polarization tensor corresponding to $\lambda$. The solution of Eq.~(\ref{II.2}) can thus be obtained using the Green's function method:
\m
h_\lambda(\boldsymbol{k},\eta) &=& 
\int_0^\eta \mathrm{d}\eta' \frac{\sin{(k(\eta-\eta'))}}{k} \frac{a(\eta')}{a(\eta)} \mathcal{S}_{\lambda} (\boldsymbol{k},\eta').\label{II.4}
\n

An observational quantity that characterize the SIGW is the energy spectrum, which is defined as GW energy density per logarithm wavelength and normalized by the critical energy $\rho_c$:
\m
\ogw(\boldsymbol{k},\eta)=\frac{1}{\rho_c} \frac{\mathrm{d}\rhoGW}{\mathrm{d}\ln k} \approx\frac{k^5}{96\pi^2\mathcal{H}^2}
\sum_\lambda \overline{ \langle h_\lambda(\boldsymbol{k},\eta) h_\lambda(\boldsymbol{-k},\eta) \rangle}.
\label{II.5} \n
The overline in Eq.~(\ref{II.5}) stands for oscillating average on $\eta$. 
The energy spectrum of SIGW can be reduced to that of primordial power spectrum of $\phi$ such that:
\m
\ogw(\boldsymbol{k},\eta) &=&
\frac{1}{6} \int_0^{+\infty} \mathrm{d}v \int_{|1-v|}^{1+v} \mathrm{d}u \frac{v^2}{u^2} \left[ 1-\left(\frac{1+v^2-u^2}{2v} \right)^2\right]^2 \mathcal{P_\phi}(u k) \mathcal{P_\phi}(v k) \overline{ I^2(u,v,k\eta)},\label{II.6}\\
I(u,v,x) &\equiv& \frac{1}{2}
\int_0^x \mathrm{d} \tilde{x} ~ \tilde{x} \sin{(x-\tilde{x})} (F(u,v,\tilde{x})+F(v,u,\tilde{x}) ). \label{II.7}
\n
The transformation has been performed in the integral that $u=p/k,~v=|\boldsymbol{p}-\boldsymbol{k}|/k$. 
Here the dimensionless power spectrum of scalar perturbation $\mathcal{P}_\phi(\boldsymbol{k})$ is defined as:
\e
\langle \phi_{\boldsymbol{k}}\phi_{\boldsymbol{k}'} \rangle \equiv \frac{2\pi^2}{k^3}\mathcal{P}_\phi(\boldsymbol{k})\delta(\boldsymbol{k}+ \boldsymbol{k}').
\q
Furthermore, we have neglected the connected four-point correlation function, which vanishes if the distribution of $\phi$ is Gaussian. For contributions from the non-Gaussianities, see e.g. \cite{Yuan:2023ofl,Li:2023xtl,Perna:2024ehx}.
The integral $\overline{I^2(u,v,k\eta)}$ in Eq.~(\ref{II.7}) was calculated in \cite{Espinosa:2018eve,Kohri:2018awv} with the limit $k\eta\rightarrow+\infty$ in which the scalar perturbations have almost decayed to zero. Consequently, $\ogw$ in the RD era can be written as
\m
\ogw(k) &=& \int_0^{+\infty} \mathrm{d}v \int_{|1-v|}^{1+v} \mathrm{d}u \mathcal{T}(u,v)\mathcal{P_R}(uk)\mathcal{P_R}(vk),
\label{ouv}\\
\mathcal{T}(u,v) &=& \frac{3}{1024 u^8 v^8} \left(u^2+v^2-3\right)^4 \left(1-(u-v)^2\right)^2
\left(1-(u+v)^2\right)^2\notag\\&&
\Bigg( \pi^2\Theta \left(u+v-\sqrt{3}\right)+
\left(\ln\frac{3-(u-v)^2}{|3-(u+v)^2|}+ \frac{4u v}{u^2+v^2-3}\right)^2 
\Bigg).\label{tuv}
\n
We have converted the scalar perturbation to the comoving curvature perturbation $\mathcal{R}=(3/2)\phi$. 
In this study, we neglect the QCD effects where the equation of state and the sound speed slightly decrease from $1/3$ during the RD era, as discussed in \cite{Abe:2020sqb}. 

In the following section, we will show a detailed calculation based on Eq.~(\ref{ouv}) in the circumstance that $\mathcal{P_R}$ has a broken power-law feature.

\section{Broken Power-law Feature}\label{III}
In this section, the curvature perturbation with a broken power-law feature in the power spectrum will be considered:
\e
\mathcal{P}_{\mathrm{PL}}(k) = A \frac{\alpha+\beta} {\beta(k/k_\ast)^{-\alpha}+\alpha(k/k_\ast)^\beta}.
\label{PPL}\q
The spectrum has a single peak at $k=k_\ast$ and is normalized by $\mathcal{P}_{\mathrm{PL}}(k_\ast) = A$. In the following sections, we use the dimensionless variable $\kappa=k/k_\ast$ instead, and refer to $\mathcal{P_R}(\kappa)$ as $\mathcal{P}_{\mathrm{PL}}(k)$ for simplicity if no ambiguity would happen.

The broken power-law peak is typically derived from single-field inflation and the parameter is further restricted by $\alpha<4$ \cite{Ballesteros:2017fsr,Byrnes:2018txb,Carrilho:2019oqg,Atal:2021jyo}. 
However, the coefficient of near-peak expansion is independent up to the $(k-k_\ast)^3$ order, which means that the broken power-law peak is always a reasonable near-peak approximation even if it is highly asymmetric. Therefore, we neglect the restriction on the parameter to step further to a general situation.

When calculating $\ogw(k)$, we will use the approximation of $\mathcal{P_R}^2(\kappa)$, namely
\m&&
\mathcal{P_R}^2(\kappa) \approx A^2
\begin{cases}
(1+\lambda)^2 \kappa^{2\alpha}
\left( 1+ d_1 \kappa^{\beta+\alpha}
\right)^2,
& \text{for}~ \kappa < \kappa_-;\\
1 + d_2 \left( \kappa^\beta-1 \right)^2,
& \text{for}~ \kappa_- < \kappa < 1;\\
1 + d_3 \left( \kappa^{-\alpha}-1\right)^2
,& \text{for}~  1 < \kappa < \kappa_+; \\
(1+\lambda^{-1})^2 \kappa^{-2\beta}
\left(1 + d_4 \kappa^ {-\alpha-\beta}
\right)^2
,& \text{for}~ \kappa > \kappa_+.
\end{cases}\label{PR2}
\n
Here $\lambda=\alpha/\beta$. The coefficients $d_m$ are listed in Eq.~(\ref{PR2coeff}). We also introduce two quantities $\kappa_-=\left(1+\lambda \right)^{-1/\alpha}$ and $\kappa_+=\left(1+\lambda^{-1}\right)^{1/\beta}$ that roughly represent the boundary of the IR and UV power-law tails of the power spectrum.
The explanation for Eq.~(\ref{PR2}) as well as the discussion on the features of power spectrum is presented in Appendix \ref{A.BPL}. The segmentation enables us to take the features of the power spectrum apart, revealing how the feature itself influences the SIGW spectrum.

\subsection{Analytical approach to SIGW} \label{III.A}

In this part, we will calculate the SIGW spectrum derived in Sec. \ref{II}, with the scalar spectrum $\mathcal{P_R}$ having a broken power-law feature described by Eq.~(\ref{PPL}).
To make use of the symmetry, we perform the transformation $x=\frac{1}{\sqrt{2}}(u+v),~ y=\frac{1}{\sqrt{2}}(u-v)$ in the integral in Eq.~(\ref{ouv}):
\m
\ogw(k) &=& \int_\frac{1}{\sqrt{2}}^{+\infty} \mathrm{d}x \int_{-\frac{1}{\sqrt{2}}}^{\frac{1}{\sqrt{2}}}\mathrm{d}y
\left(1-2y^2\right)^2 F(x,y)
\mathcal{P_R}\left[q\left(1+\frac{y}{x}\right)\right] \mathcal{P_R}\left[q\left(1-\frac{y}{x}\right)\right],\label{Oxy}
\\
F(x,y) &=&
\frac{ 3\left(1-2x^2\right)^2 \left(x^2+y^2-3\right)^4 }{4 (x^2-y^2)^8}
\left[\pi^2 \Theta \left(x-\frac{\sqrt{6}}{2}\right) +\left( \ln \frac{3-2y^2}{|3-2x^2|}+\frac{2(x^2-y^2)} {x^2+y^2-3}\right)^2\right],\label{Fxy}
\n
Here we have defined $q=\kappa x/\sqrt{2}$ for simplicity. Obviously, most of the factors of $F(x,y)$ are not sensitive to $y$ since $y^2<1/2$, except for the region near the zero point of $(x^2+y^2-3)$ and $(x^2-y^2)$.
The former contributes little deviation to $\ogw$ since it is suppressed by the integration on the nearby peak if $\mathcal{P_R}$ has a width $\gtrsim 10^{-1}$, while below that the leading order of $y$ is much smaller than that of $x$.
Therefore, we consider the region $x>\sqrt{6}/2$ first and deal with the integration on $y$ separately:
\begin{align}
&\quad\ogw^{x>\frac{\sqrt{6}}{2}}(k) = \int_\frac{\sqrt{6}}{2}^{+\infty} \mathrm{d}x ~ F(x,0)\mathcal{P}_\mathcal{R}^2(q) I_y(x,q)
,\label{III.B.1}\\
I_y(x,q) =&
\int_{-\frac{1}{\sqrt{2}}}^{\frac{1}{\sqrt{2}}} \mathrm{d}y \left(1-2y^2\right)^2 \frac{F(x,y)}{F(x,0)}
\frac{\mathcal{P_R}\left[q\left(1+\frac{y}{x}\right)\right]\mathcal{P_R}\left[q\left(1-\frac{y}{x}\right)\right]}{\mathcal{P}_{\mathcal{R}}^{2}(q)} ,\notag\\
\approx& \int_{-\frac{1}{\sqrt{2}}}^{\frac{1}{\sqrt{2}}} \mathrm{d}y \left(1-2y^2\right)^2 (1+4~\frac{y^2}{x^2})
\frac{\mathcal{P_R}\left[q\left(1+\frac{y}{x}\right)\right]\mathcal{P_R}\left[q\left(1-\frac{y}{x}\right)\right]}{\mathcal{P}_{\mathcal{R}}^{2}(q)} .\label{III.B.2}    
\end{align}
In Eq.~(\ref{III.B.2}), we connect the asymptotic behavior of $F(x,y)/F(x,0)$ at $x=\sqrt{6}/2$ and $x\rightarrow+\infty$ up to $y^2$ order, where the $x^{-2}$ dependence is selected for further convenience.
The analytical expression of $I_y(x,q)$ and $\ogw^{x>\frac{\sqrt{6}}{2}}(k)$ are studied in Appendix~\ref{A.B}, and we give the result here directly:
\begin{subequations}\m
\ogw^{x>\frac{\sqrt{6}}{2}}(k) &=&
\ogw^{q<\kappa_-}(k)+\ogw^{\kappa_-<q<1}(k)+
\ogw^{1<q<\kappa_+}(k)+\ogw^{q>\kappa_+}(k),\label{ogw1}\\
\ogw^{q<\kappa_-}(k) &\approx&  A^2(1+\lambda)^2
\sum_{\rho} \sum_{j=0}^2 a_{\rho} C_2^j d_1^j
\left(\frac{\kappa}{\sqrt{2}}\right)^ {2\alpha+j(\alpha+\beta)}
\Omega(\rho+2\alpha+j(\alpha+\beta),x)\Big{|}_{\sqrt{6}/2}^{R_1}, \label{ogw1.1}\\
\ogw^{\kappa_-<q<1}(k) &\approx& A^2
\sum_{\rho} b_{\rho}\left(
\Omega(\rho,x)+d_2 \sum_{j=0}^2 C_2^j (-1)^j 
\left(\frac{\kappa}{\sqrt{2}}\right)^{j\beta}
\Omega(\rho+j\beta,x)\right) \Bigg{|}_{R_1}^{R_2},\label{ogw1.2}\\
\ogw^{1<q<\kappa_+}(k) &\approx& A^2
\sum_{\rho} b_{\rho}\left(
\Omega(\rho,x)+d_3 \sum_{j=0}^2 C_2^j (-1)^j  
\left(\frac{\kappa}{\sqrt{2}}\right)^{-j\alpha}
\Omega(\rho-j \alpha,x)
\right) \Bigg{|}_{R_2}^{R_3},\label{ogw1.3}\\
\ogw^{q>\kappa_+}(k) &\approx&  A^2(1+\lambda^{-1})^2
\sum_{\rho} \sum_{j=0}^2 c_{\rho} C_2^j d_4^j
\left(\frac{\kappa}{\sqrt{2}}\right)^ {-2\beta-j(\alpha+\beta)}
\Omega(\rho-2\beta-j(\alpha+\beta),x)
\Big{|}_{R_3}^{+\infty}\label{ogw1.4}
.\n\end{subequations}
The segmentation in Eq.~(\ref{ogw1}) is based on Eq.~(\ref{PR2}). The sum on $\rho$ in Eq.~(\ref{ogw1.1})-(\ref{ogw1.4}) is introduced during the simplification of $I_y(x,q)$, and the summation rule is different for different cases, which is listed in Eq.~(\ref{brho})(\ref{crho}) and (\ref{arho}). For example according to Eq.~(\ref{brho}), we have
\begin{align}
    \sum_{\rho} b_{\rho}\Omega(\rho,x)= b_{0}\Omega(0,x) + b_{-2}\Omega(-2,x) = \frac{8\sqrt{2}}{15}\Omega(0,x) + \frac{8\sqrt{2}}{15} \frac{1}{14}(4-\alpha\beta) \Omega(-2,x) .
\end{align}
Besides, all the defined coefficients used in Eq.~(\ref{ogw1.1})-(\ref{ogw1.4}) are listed below Eq.~(\ref{Ogw1.4}) in Appendix~\ref{A.B}.
The function $\Omega(\rho',x)$ denotes the integration of $F(x,0)x^{\rho'}$, which can be done directly and the analytical results are listed in Appendix \ref{A.Orx}.

As for the region $x<\sqrt{6}/2$, $F(x,0)$ decays to zero rapidly as $x$ decreases, and hence the contribution is significant only if $\kappa>2/\sqrt{3}$ and the peak of $\mathcal{P_R}^2(q)$ enters this region. 
Since the feature of this region is quite simple, we will take the leading order of $x\rightarrow(\sqrt{6}/2)_-$, $v\rightarrow0$ and $x\rightarrow\sqrt{2}/2$ respectively for different areas and add them up.

Firstly for $x\rightarrow(\sqrt{6}/2)_-$, we take the expansion around $(x-\sqrt{6}/2)$ to the zeroth order and $y$ to the second order:
\m
F(x,y) \approx \frac{16}{27} (1+\frac{8}{3}y^2)
\ln^2 \left[\frac{4 e^2}{\sqrt{6}}
\left( x-\frac{\sqrt{6}}{2} \right)
\right].
\n
We perform the integration from the zero point of the expansion at $x=(\sqrt{6}/2)(1-(2e^2)^{-1})$. 
The width on $x$ direction is $<10^{-1}$, and the integrand is greatly enhanced nearby the singularity at $x=\sqrt{6}/2$.
As a consequence, the power spectrum in Eq.~(\ref{Oxy}) can be treated as constant of $x$ in this region, and the contribution from this part can thus be obtained:
\m
\ogw^{\frac{\sqrt{6}}{2}_-}(k) 
&\approx& \frac{16}{27} \mathcal{P_R}^2\(\frac{\sqrt{3}}{2}\kappa\) I_y\(\frac{\sqrt{6}}{2},\frac{\sqrt{3}}{2}\kappa\) \int_{\frac{\sqrt{6}}{2}(1-\frac{1}{2e^2})}^{\frac{\sqrt{6}}{2}} \mathrm{d}x \ln^2 \left[\frac{4 e^2}{\sqrt{6}} \left( x-\frac{\sqrt{6}}{2} \right) \right] ,
\notag\\&=& 
\frac{8\sqrt{6}}{27e^2} \mathcal{P_R}^2\(\frac{\sqrt{3}}{2}\kappa\)
I_y\(\frac{\sqrt{6}}{2},\frac{\sqrt{3}}{2}\kappa\)
.\label{ogw2.1}\n
Here $I_y(\sqrt{6}/2,\sqrt{3}\kappa/2)$ is defined in Eq.~(\ref{III.B.2}) and we give the analytical expression in Eq.~(\ref{IyNormal}).

Then in the region $v<(\sqrt{3}-1)/2$, we take the leading order of Eq.~(\ref{ouv}) with $v\rightarrow0$:
\m
\ogw^{v\rightarrow0}(k) &\approx&  \frac{1}{3}\mathcal{P_R}(\kappa)\int_0^{\frac{\sqrt{3}-1}{2}} \mathrm{d}v~
v^3\mathcal{P_R}(v \kappa) 
\int_{-1}^1 \mathrm{d}\delta~ (\delta^2-1)^2,
\\&=& 
\frac{32(\alpha+\beta)} {45 \beta(\alpha +4)  \left(\sqrt{3}+1\right)^{\alpha+4}}
A\mathcal{P_R}(\kappa) \kappa^{\alpha} 
\,_2F_1\left(1, \frac{\alpha+4}{\alpha+\beta}; \frac{\alpha+4}{\alpha+\beta}+1; -\frac{\lambda \kappa^{\alpha+\beta}} {(\sqrt{3}+1)^{\alpha+\beta}}\right)
,\label{ogw2.2}
\n
where $\delta$ is defined by $u=1+v\delta$, $_2 F_1$ is the hyper-geometric function. The result has been doubled considering the identical region $u<(\sqrt{3}-1)/2$. 

Finally, we take the leading order of $F(x,y)$ in Eq.~(\ref{Fxy}) at $x=1/\sqrt{2}$. We use the same method as Eq.~(\ref{III.B.1}) and the remaining integral reduces to
\m
&&\ogw^{\frac{\sqrt{2}}{2}_+}(k) \approx 2400(2+5\ln(2/3))^2
\int_{ \frac{1}{\sqrt{2}} }^{ \frac{\sqrt{6}}{2} } \mathrm{d}x ~ \left(x-\frac{1}{\sqrt{2}}\right)^2  \mathcal{P_R}^2(q)
\int_{\frac{\sqrt{6}-\sqrt{2}}{2}-x}^{\frac{\sqrt{2}-\sqrt{6}}{2}+x} \mathrm{d}y \frac{\mathcal{P_R}\left[q\left(1+\frac{y}{x}\right)\right]\mathcal{P_R}\left[q\left(1-\frac{y}{x}\right)\right]}{\mathcal{P}_{\mathcal{R}}^{2}(q)}.
\n
When integrating $y$, we regard the integrand as constant if $\alpha\beta\leq3$, which means the peak width on $y$ direction do not enter this region. As for the case $\alpha\beta>3$, we focus on the near-peak approximation near $q=1$ and the integration on $y$ can be approximated to
\begin{equation}
\int_{\frac{\sqrt{6}-\sqrt{2}}{2}-x}^{\frac{\sqrt{2}-\sqrt{6}}{2}+x} \mathrm{d}y \frac{\mathcal{P_R}\left[q\left(1+\frac{y}{x}\right)\right]\mathcal{P_R}\left[q\left(1-\frac{y}{x}\right)\right]}{\mathcal{P}_{\mathcal{R}}^{2}(q)} \approx 2x\int_0^{(\alpha\beta)^{-1/2}}\mathrm{d}s \exp{-\alpha\beta s^2}=\frac{\sqrt{\pi } \text{erf}(1) x}{\sqrt{\alpha \beta }},
\end{equation}
where $s=y/x$. Therefore, the result of $\ogw$ in this region is that:
\begin{gather}
\ogw^{\frac{\sqrt{2}}{2}_+}(k) \approx C_\kappa \sum_{m=0}^2 \left(-\sqrt{2}\right)^m C_2^m 
\times \begin{cases}
\left(2 I_1\left(2\alpha+m+1,x\right)+
(\sqrt{2}-\sqrt{6})
I_1\left(2\alpha+m,x\right)
\right)\Big|_{\sqrt{2}/2}^{\sqrt{6}/2}
& \text{for}~\alpha\beta\leq3,\\
\dfrac{\sqrt{\pi } \text{erf}(1)}{\sqrt{\alpha \beta }} I_1\left(2\alpha+m+1,x\right) \Big|_{\sqrt{2}/2} ^{\sqrt{6}/2} & \text{for}~\alpha\beta>3.
\end{cases}\label{ogw2.3}\\
C_\kappa = 1200\(2+5\ln\frac{2}{3}\)^2 A^2 2^{-\alpha } (1+\lambda)^2 \kappa^{2\alpha} ,\quad
I_1(r,x)
=\frac{x^{r+1}}{r+1} \,_2F_1\left(2, \frac{r+1}{\alpha +\beta};\frac{r+1}{\alpha +\beta}+1;-\lambda\,q^{\alpha+\beta}\right). 
\end{gather}

\begin{figure}
\centering
    \begin{minipage}{0.49\linewidth}
      \centering
      \includegraphics[width=0.99\linewidth]{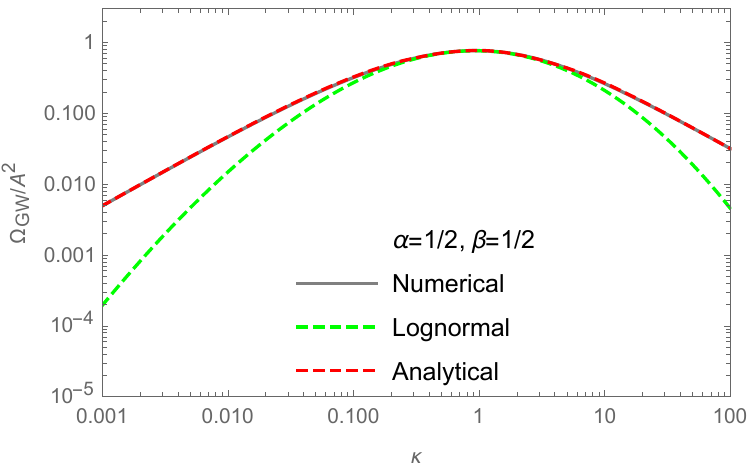}
      \label{figogw1}
    \end{minipage}
    \begin{minipage}{0.49\linewidth}
      \centering
      \includegraphics[width=0.99\linewidth]{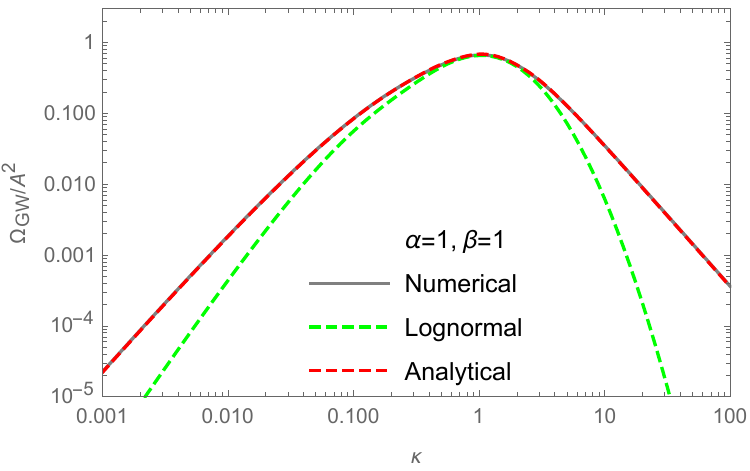}
      \label{figogw2}
    \end{minipage}\\
    \begin{minipage}{0.49\linewidth}
      \centering
      \includegraphics[width=0.99\linewidth]{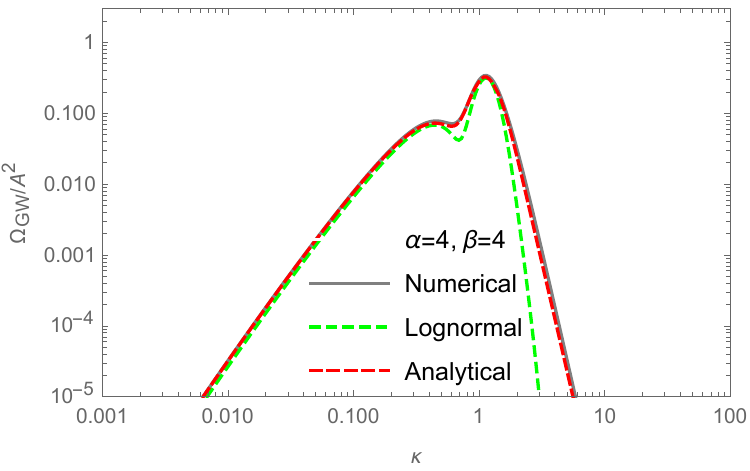}
      \label{figogw3}
    \end{minipage}
    \begin{minipage}{0.49\linewidth}
      \centering
      \includegraphics[width=0.99\linewidth]{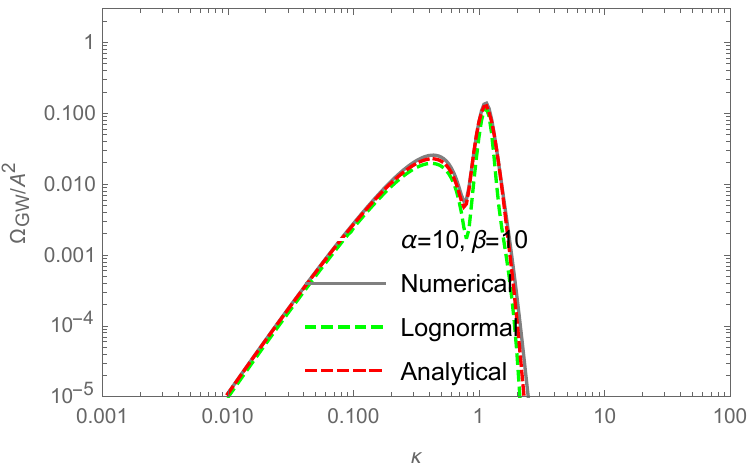}
      \label{figogw4}
    \end{minipage}\\
    \begin{minipage}{0.49\linewidth}
      \centering
      \includegraphics[width=0.99\linewidth]{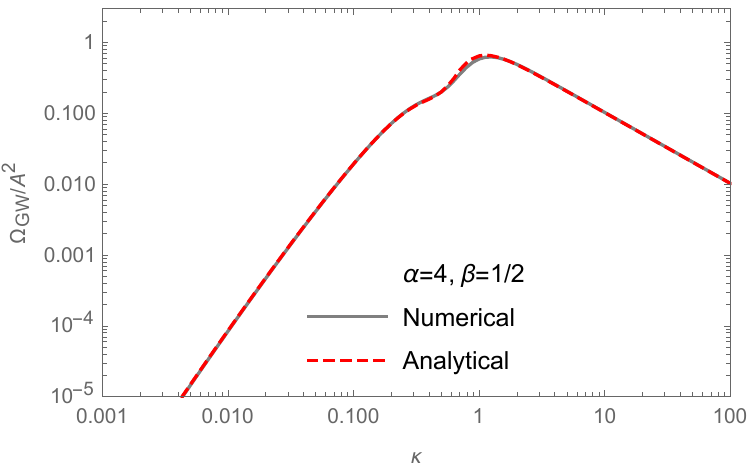}
      \label{figogw5}
    \end{minipage}
    \begin{minipage}{0.49\linewidth}
      \centering
      \includegraphics[width=0.99\linewidth]{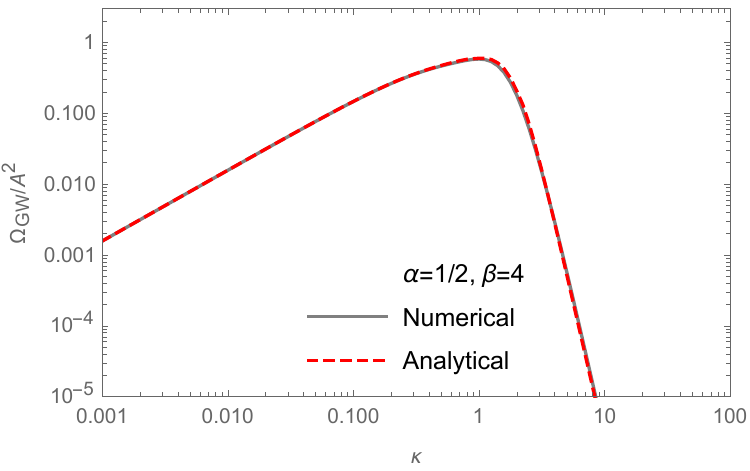}
      \label{figogw6}
    \end{minipage}
    \caption{The analytical results of the normalized $\ogw$ (where we divide it by $A^2$) generated by broken power-law spectrum as a function of $\kappa=k/k_*$ together with the numeric results for different values of $\alpha$ and $\beta$. We also show the normalized $\ogw$ generated by log-normal power spectrum which is characterized in Eq.~(\ref{LN}) with $\Delta=\alpha^{-1}$.}
    \label{figogwI}
\end{figure}

The final result of $\ogw(k)$ induced by a broken power-law power spectrum is the sum of Eqs. (\ref{ogw2.1}) (\ref{ogw2.2}) (\ref{ogw2.3}) and (\ref{ogw1}):
\begin{mdframed}[linecolor=black, linewidth=1pt, backgroundcolor=white, nobreak=true]
\textbf{Final result:}
\begin{equation}
\ogw(k) \approx 
\ogw^{x>\frac{\sqrt{6}}{2}}(k)+\ogw^{\frac{\sqrt{6}}{2}_-}(k)+ \ogw^{v\rightarrow0}(k)+\ogw^{\frac{\sqrt{2}}{2}_+}(k)
\label{ogw}.
\end{equation}
$\qquad$
\end{mdframed}
Eq.~(\ref{ogw}) is our main result and readers can refer back to the intermediate results as needed. 
The relevant parameters are listed in the appendix for a quick approach.
The analytical result is compared to the numerical result in Fig. \ref{figogwI}. 
We also present a code to reproduce our results \cite{chen2024github}.

\subsection{IR asymptotic behaviors}\label{III.B}

Furthermore, it is also of vital importance to work out the asymptotic behavior of $\ogw(k)$ to show a straightforward feature. 
Several works have found the simplicity of $\ogw(k)$ in IR region for different models\cite{Yuan:2019wwo,Pi:2020otn,Yi:2023mbm}, and the distinguishing feature will be useful to search for SIGWs.
In this part, we show the IR behaviors of our result in Sec. \ref{III.A}.
We only show the dominant term here, with the exact result and detailed calculation given in Appendix~\ref{A.IR}. 
In the results below, we put the $\kappa$-dependent factors into square brackets to make them more intuitive.

Returning to our result of $\ogw$ in \Eq{ogw}, we find that in the IR region, i.e. $\kappa\ll\kappa_-$, the main contribution is from the first term $\ogw^{x>\frac{\sqrt{6}}{2}}(k)$ which is defined in Eq.~(\ref{III.B.1}). The reason is that other terms correspond to the region $\sqrt{2}/2<x<\sqrt{6}/2$, where the power spectrum is suppressed by $x^{2\alpha}$ and $F(x,y)$ is suppressed by $(1-2x^2)^2$. 
In this case, we split Eq.~(\ref{III.B.1}) into two parts:
\begin{equation}
\ogw(k)\approx\ogw^{x>\frac{\sqrt{6}}{2}}(k) =\int_{\frac{\sqrt{6}}{2}}^{\frac{\sqrt{2}\kappa_-}{\kappa}} \mathrm{d}x ~ F(x,0)\mathcal{P}_\mathcal{R}^2(q) I_y(x,q)+\int_{\frac{\sqrt{2}\kappa_-}{\kappa}}^{+\infty} \mathrm{d}x ~ F(x,0)\mathcal{P}_\mathcal{R}^2(q) I_y(x,q),\label{ogwIR}
\end{equation}
where the first integral refers to the $\ogw^{q<\kappa_-}(k)$ term in Eq.~(\ref{ogw1.1}). Since $x>\sqrt{2}\kappa_-/\kappa\gg1$ in the second integral, we have
\begin{equation}
F(x,y) \approx \frac{3}{x^4}\left(\pi^2+ \ln^2\frac{2x^2}{3e^2}\right),\quad
I_y(x,q) \approx \int_{-\frac{1}{\sqrt{2}}}^{\frac{1}{\sqrt{2}}} \mathrm{d}y \left(1-2y^2\right)^2 \frac{\mathcal{P_R}\left[q\left(1+\frac{y}{x}\right)\right]\mathcal{P_R}\left[q\left(1-\frac{y}{x}\right)\right]}{\mathcal{P}_{\mathcal{R}}^{2}(q)}.
\label{FxyIR}\end{equation}
In this limit, $F(x,y)$ is no longer relevant to $y$.
Using the approximation on $F(x,y)$, we can reduce $\Omega(\rho,x)$, which we have used to describe our result Eq.~(\ref{ogw1})-(\ref{ogw1.4}), to a simple form of
\m
\Omega(\rho,x)=\int\mathrm{d}x ~ x^\rho F(x,0)
&\approx&
\begin{cases}
\dfrac{12}{\rho-3}x^{\rho-3}\ln^2 x
&\text{for}~\rho\neq3,\\\\
4\ln^3 x
&\text{for}~\rho=3.
\end{cases}\label{ogwrhoinf1}
\n
We only keep the leading order of $\ln x$ here, while the complete result for the $x^{\rho-3}$ order is in Eq.~(\ref{ogwrhoinf}). The approximation above can greatly reduce the complexity of Eqs.(\ref{ogw1})-(\ref{ogw1.4}).

Next, we will simplify $\Omega_{\mathrm{GW}}(k)$ one step futher. Given that the following calculation depends on the value of $\alpha$ and $\beta$, we will split them into different cases, namely $\alpha<3/2,~\alpha>3/2,~\alpha=3/2$ and $\alpha, \beta\gg30$.

Firstly for $\alpha<3/2$, the enhancement in the power spectrum is too slow to cover the decay of $F(x,y)$ in Eq.~(\ref{FxyIR}), and therefore the second integral of Eq.~(\ref{ogwIR}) is suppressed by the $x^{-4}$ factor of $F(x,y)$.
In this case, the dominant term is the first integral in Eq.~(\ref{ogwIR}), i.e. the $\ogw^{q<\kappa_-}(k)$ term in Eq.(\ref{ogw1}), and the dominant term is that
\begin{equation}
\ogw\approx\ogw^{q<\kappa_-}(k) \approx  -A^2 (1+\lambda)^2 \[ \left(\frac{\kappa}{\sqrt{2}}\right)^ {2\alpha}\] \sum_{\rho=0,-2} a_{\rho}\Omega(\rho+2\alpha,\frac{\sqrt{6}}{2}).\label{ogwIRA}
\end{equation}
The sum of $a_{\rho}\Omega(\rho+2\alpha,\sqrt{6}/2)$ is constant in IR region, and therefore, $\ogw^{q<\kappa_-}(k)$ behaves as $\kappa^{2\alpha}$. 
The $\kappa^{2\alpha}$ behavior is natural. As $\kappa\rightarrow0$, the integrand in $x\gg1$ region decays with the power index $2\alpha-4<-1$ and the integral is no longer relevant to the upper limit $(\sqrt{2}\kappa_-/\kappa)$. As a result, the remaining $\kappa$-dependent factor in the integral is $\kappa^{2\alpha}$ and thus we have $\Omega_{\mathrm{GW}}(k)\propto k^{2\alpha}$ in the IR region.
 
Then for $\alpha>3/2$, the enhancement in the power spectrum is large enough so that the main contribution comes from the near-peak region of the power spectrum.
In this case, the IR behavior in the limit of $\Delta_{\kappa,\mathrm{IR}}\gg1$ and $\Delta_{\kappa,\mathrm{IR}}\ll1$ is different, where $\Delta_{\kappa,\mathrm{IR}}$ is defined by
\e
\Delta_{\kappa,\mathrm{IR}} = \frac{225}{16}\frac{1}{\kappa^2\alpha\beta}.\notag
\q
This parameter is roughly the squared ratio of the power spectrum width to $\kappa$, and determines the behavior of $I_y(x,q)$ in Eq.~(\ref{FxyIR}).
Note that since $\kappa\ll\kappa_-<1$, the limit $\Delta_{\kappa,\mathrm{IR}}\ll1$ means that $\alpha\beta$ is extremely large.
We show the corresponding results below separately.

In the case $\Delta_{\kappa,\mathrm{IR}}\gg1$, in other words $\kappa\ll(15/4)(\alpha\beta)^{-1/2}$, $I_y(x,q)$ reduces to a constant near the peak of power spectrum. This means that the coefficients $a_\rho,~b_\rho$ and $c_\rho$ used in Eq.~(\ref{ogw1.1})-(\ref{ogw1.4}) are also constant, and therefore, their dominant terms have the same feature that
\m
\left(\frac{\kappa}{\sqrt{2}}\right)^\rho \Omega(\rho,x) \Bigg{|}_{\frac{\sqrt{2}}{\kappa}\kappa_i} ^{\frac{\sqrt{2}}{\kappa}\kappa_j} \approx 3\sqrt{2} \frac{\kappa_j^{\rho-3}-\kappa_i^{\rho-3}}{\rho-3}  \[\kappa^3\ln^2\kappa\].
\label{k3ln2k1}\n
Since the $\kappa$-dependent factor $\kappa^3\ln^2\kappa$ is independent of $\rho$, the dominant term of $\ogw$ has the same feature that
\begin{equation}
\ogw(k)\approx \ogw^{q<\kappa_-}(k)+ \ogw^{\kappa_-<q<1}(k) + \ogw^{1<q<\kappa_+}(k)+\ogw^{q>\kappa_+}(k)\propto\kappa^3\ln^2\kappa.\label{ogwIRB}
\end{equation}

In the critical case $\alpha=3/2$, we find the result of $\ogw^{q<\kappa_-}(k)$ is different considering the different $\ln^3x$ feature in Eq.~(\ref{ogwrhoinf1}):
\m
\ogw^{q<\kappa_-}(k) 
&\approx&
\frac{16}{15} A^2 (1+\lambda)^2 \[-\kappa^3 \ln^3\kappa\].\label{ogwIRC}
\n
Here the dominant term turns into $\kappa^{3}\ln^3\kappa$. 


Finally for $\Delta_{\kappa,\mathrm{IR}}\ll1$, in other words $(15/4)(\alpha\beta)^{-1/2}\ll\kappa\ll\kappa_-$, $I_y(x,q)$ is proportional to $\kappa^{-1}$ near the peak, so that $\ogw$ also has an extra $\kappa^{-1}$ factor. Besides, the remaining $\kappa$-dependent factor in Eq.~(\ref{ogwIR}) is also Eq.~(\ref{k3ln2k1}) in the dominant term, so that $\ogw$ has the feature that
\begin{equation}
\ogw(k)\approx \ogw^{q<\kappa_-}(k)+ \ogw^{\kappa_-<q<1}(k) + \ogw^{1<q<\kappa_+}(k)+\ogw^{q>\kappa_+}(k)\propto\kappa^2\ln^2\kappa.\label{ogwIRD}
\end{equation}
This feature appears only if both $\alpha$ and $\beta$ are extremely large, or explicitly $\alpha,\beta\gg30$.

To summarize, different IR behaviors of $\ogw$ are listed in Table \ref{tab:IR}.
As $\alpha$ increases, $\kappa$-dependence changes. The transition $\kappa^{2\alpha}\to\kappa^3\ln^3\kappa\to\kappa^3\ln^2\kappa$ is a typical feature induced by the power-law IR tail of the scalar power spectrum, and these results recover the previous results in \cite{Xu:2019bdp,Atal:2021jyo,Balaji:2022dbi,Yi:2023mbm} where the power spectrum with a power-law IR tail is also considered.
Then if both $\alpha$ and $\beta$ are extremely large, the broken power-law spectrum is highly peaked and the transition to $\kappa^2\ln^2\kappa$ emerges.
The transition $\kappa^3\ln^2\kappa\to\kappa^2\ln^2\kappa$ is a typical feature induced by a peaked power spectrum, and is also found in \cite{Yuan:2019wwo,Pi:2020otn} where they used a log-normal power spectrum and a general peaked spectrum with cut-off on both sides. We emphasize that our method is more accurate since we give the full result of $\kappa^3$ and $\kappa^2$ order in Appendix~\ref{A.IR} instead of only the scaling.

It should be noticed that the method in this section can also be extended to more general cases. 
On the one hand, Eq.~(\ref{k3ln2k1}), or the exact expression in Eq.~(\ref{k3ln2k}), can be treated as part of a general $\ogw$.
Looking back to the second integral in Eq.~(\ref{ogwIR}) within the range $\sqrt{2}\kappa_i/\kappa<x<\sqrt{2}\kappa_j/\kappa$, it reduce to Eq.~(\ref{k3ln2k1}) naturally if the power spectrum can be treated as power-law therein, and $I_y(x,q)$ can be treated as constant of $x$ locally.
The conditions above are always true.
If the width of power spectrum is much larger than $\kappa$, we have $I_y(x,q)\approx8\sqrt{2}/15$, so that part of $\ogw$ has the feature $\kappa^3\ln^2\kappa$ naturally in this case.
If the width of power spectrum is much smaller than $\kappa$, it is also universal that 
\begin{equation}
    I_{y}(\sqrt{2}/\kappa,1) \approx \int_{-\frac{1}{\sqrt{2}}}^{\frac{1}{\sqrt{2}}} \mathcal{P_R}\left(1+\frac{\kappa y}{\sqrt{2}}\right) \mathcal{P_R}\left(1-\frac{\kappa y}{\sqrt{2}}\right) \mathrm{d}y
    \approx \frac{\sqrt{2}}{\kappa} \int_0^{+\infty}\mathcal{P_R}(t)\mathcal{P_R}(t^{-1})\mathrm{d}\ln t
    \propto\kappa^{-1}.
\end{equation}
Here $t=\exp(\kappa y/\sqrt{2})$. This is the value of $I_y(x,q)$ near the peak of $\mathcal{P_R}(q)$, and as a result, $\ogw$ has a feature $\kappa^2\ln^2\kappa$ inevitably in this case.

On the other hand, the discussion on the first integral of Eq.~(\ref{ogwIR}), i.e. the $\ogw^{q<\kappa_-}(k)$ term, is applicable to all the cases in which the power spectrum has a power-law IR tail. As a result, the transition of dependence $\kappa^{2\alpha}\to\kappa^3\ln^3\kappa\to\kappa^3\ln^2\kappa$ is generic in this case.

\begin{table}
    \centering
    \begin{tabular}{c|ccc}
    \hline\hline
    Model   & Range of $\kappa$ & IR Behavior\\
    \hline
    $\alpha<3/2$  & $\kappa\ll\kappa_-$  & $\kappa^{2\alpha}$\\
    $\alpha=3/2$  & $\kappa\ll\kappa_-$  & $\kappa^{3}\ln^3\kappa$\\
    $\alpha>3/2$  & $\kappa\ll\kappa_-,~ \kappa\ll(15/4)(\alpha\beta)^{-1/2}$  & $\kappa^{3}\ln^2\kappa$ &\\
    $\alpha,\beta\gg30$  & $(15/4)(\alpha\beta)^{-1/2}\ll\kappa\ll\kappa_-$ & $\kappa^{2}\ln^2\kappa$ &\\
    \hline\hline
    \end{tabular}
    \caption{Different IR behavior in different parameter ranges. Properties of the dominant terms are given in Eq.~(\ref{k3ln2k1}), where $\kappa_i$ includes $\kappa_-,~1,~\kappa_+$ and $+\infty$.}
    \label{tab:IR}
\end{table}

\subsection{UV asymptotic behavior}\label{III.C}

In this section, we turn into the UV behavior of $\ogw$. Different from the IR region, all the terms in Eq.~(\ref{ogw}) may have significant contribution here.
In UV region, i.e. $\kappa\gg\kappa_+$, 
most region of the integral in Eq.~({\ref{Oxy}}) reduces to
\m
A^2 (1+\lambda^{-1})^2 \(\frac{\kappa}{\sqrt{2}}\)^{-2\beta} \int \mathrm{d}x \int\mathrm{d}y
\left(1-2y^2\right)^2 F(x,y) x^{-2\beta} \(1-\frac{y^2}{x^2}\)^{-\beta},
\n
except the $ \ogw^{v\rightarrow0}(k)$ part in Eq.~(\ref{ogw}). The integral above is irrelevant to $\kappa$ so that part of $\ogw(k)$ evidently behaves as $\kappa^{-2\beta}$. This can also be obtained from our analytical result:
\m
\ogw^{x>\frac{\sqrt{6}}{2}}(k) &\approx& A^2 \kappa^{-2\beta} 2^\beta  (1+\lambda^{-1})^2
\sum_{\rho=0,-2} c_{\rho}
\Omega(\rho-2\beta,x)
\Big{|}_{\sqrt{6}/2}^{+\infty}\label{ogwUV1},\\
\ogw^{\frac{\sqrt{6}}{2}_-}(k) &\approx&  A^2 \kappa^{-2\beta}  \frac{128(1+\lambda^{-1})^2}{135\sqrt{3}e^2} \left(1+\frac{4+\beta}{21}\right) \left(\frac{3}{4}\right)^{-\beta} 
;\label{ogwUV2.1}\\
\ogw^{\frac{\sqrt{2}}{2}_+}(k) &\approx&  A^2 \kappa^{-2 \beta }
75\ 2^{\beta +5} (2+5 \ln (2/3))^2 (1 +\lambda^{-1})^2 
\sum_{m=0}^2 (-1)^m 2^{m/2} C_2^m \notag\\&&\times
\begin{cases}
\(\dfrac{x^{-2 \beta +m+2}}{-2 \beta +m+2}+ \dfrac{\sqrt{2}-\sqrt{6}}{2} \dfrac{x^{-2 \beta +m+1}}{-2 \beta +m+1}
\)\Bigg|_{\sqrt{2}/2} ^{\sqrt{6}/2} &\text{for}~\alpha\beta\leq3,\\
\dfrac{\text{erf}(1)\sqrt{\pi}}{2\sqrt{\alpha\beta}}
\dfrac{x^{-2 \beta +m+2}}{-2 \beta +m+2} \Bigg|_{\sqrt{2}/2} ^{\sqrt{6}/2} &\text{for}~ \alpha\beta>3.
\end{cases}
\label{ogwUV2.3}
\n
All the terms are proportional to $\kappa^{-2\beta}$ here, which are contributions from the UV tail of the power spectrum on $q>\kappa_+$. However, the UV behavior of $\ogw^{v\rightarrow0}(k)$ is different:
\begin{equation}
    \ogw^{v\rightarrow0}(k) \approx  A^2 \frac{32}{45\alpha} \left(\lambda^{-1}+1\right) \times
\begin{cases}
    \left[\left(\sqrt{3}+1\right)^{-4+\beta} \dfrac{\alpha+\beta}{4-\beta} \kappa^{-2\beta} +\pi \lambda ^{\frac{\beta -4}{\alpha +\beta }} \csc \left(\dfrac{\pi  (\alpha +4)}{\alpha +\beta }\right) \kappa^{-\beta -4}\right] &\text{for}~\beta\neq4,\\
    (\alpha +4) \kappa^{-8} \left(\ln\kappa -\ln\left(\sqrt{3}+1\right) + \dfrac{1}{\alpha+4} \ln\dfrac{\alpha }{4} \right) &\text{for}~\beta=4;
\end{cases}\label{ogwUV2.2}
\end{equation}
The contribution from Eq.~(\ref{ogwUV2.2}) has a different $\kappa^{-4-\beta}$ term, which is from the near-peak region of power spectrum overlapped with the UV tail.
In the critical case $\beta=4$, $\ogw$ behaves differently as $\kappa^{-8}\ln\kappa$. To summarize, the UV behavior of $\ogw$ is that:
\begin{align}
    \ogw(k) = \ogw^{x>\frac{\sqrt{6}}{2}}(k)+\ogw^{\frac{\sqrt{6}}{2}_-}(k)+ \ogw^{v\rightarrow0}(k)+\ogw^{\frac{\sqrt{2}}{2}_+}(k) \propto
\begin{cases}
    \kappa^{-2\beta} &\text{for}~\beta<4\\
    \kappa^{-8}\ln\kappa&\text{for}~\beta=4\\
    \kappa^{-4-\beta}&\text{for}~\beta>4
\end{cases}\label{ogwUV}
\end{align}
The approximation is valid if the peak of power spectrum appears only in the integral in $\ogw^{v\rightarrow0}(k)$ and this roughly means $\kappa>(\sqrt{2}+\sqrt{6})\kappa_+$.

The results in Eqs. (\ref{ogwUV1})-(\ref{ogwUV2.3}) and Eq.~(\ref{ogwUV}) can be extended to a general case of a power spectrum with a power-law UV tail.
In this case, similar feature was found in \cite{Xu:2019bdp,Atal:2021jyo,Balaji:2022dbi}.
Here, we derive the coefficients of $\kappa^{-4-\beta}$ and $\kappa^{-8}\ln\kappa$ in Eq. (44) which can be used as a good approximation of the SIGW energy spectrum.

\section{Conclusion}\label{IV}
In this paper, we have derived an analytical result for the energy spectrum $\ogw(k)$ of SIGWs induced by the scalar perturbations with a broken power-law power spectrum with high precision. These scalar perturbations can be generated from single-field inflation, and we consider the SIGWs produced during the RD era, during which PBHs may have formed and now constitute a significant fraction of dark matter.
The shape of $\ogw(k)$ is sensitive to the form of the scalar power spectrum, particularly to the enhancing and decaying power indices $\alpha$ and $\beta$ for IR and UV regions respectively, and is proportional to the square of the amplitude, $A$. The complete analytical result is given in Eq.~(\ref{ogw}). 
Moreover, our result can be used as a near-peak approximation in more general cases since the broken power-law power spectrum has abundant near-peak feature.

To facilitate the calculation, we separate the power-law tails of $\mathcal{P_R}$ and reconstruct the near-peak region as shown in Eq.~(\ref{PR2}). 
Considering the transfer function part of $\ogw(k)$, or $F(x,y)$ in Eq.~(\ref{Oxy}), we separate the integral from the resonant peak of the source term at $x=\sqrt{6}/2$, corresponding physically to $|\boldsymbol{k-p}|+p=\sqrt{3}k$. 
The region $x>\sqrt{6}/2$ is the main contribution of the integral, and contribute to the detailed shape of $\ogw$. The contribution of this part is given in Eq.~(\ref{ogw1}), where the contributions from different regions of $\mathcal{P_R}$ are given separately in Eqs. (\ref{ogw1.1})-(\ref{ogw1.4}). 
On the other region $x<\sqrt{6}/2$, the transfer function decays rapidly and the complete result is given in Eq.~(\ref{ogw2.1}), (\ref{ogw2.2}) and (\ref{ogw2.3}).

The calculation is quite different for $\alpha\beta\leq6$ and $\alpha\beta>6$, which distinguish whether the width of $\mathcal{P_R}(\kappa)$ enters the integral on $y$ direction when resonating with the mean peak of $F(x,y)$ at $x=\sqrt{6}/2$. 
In the case $\alpha\beta>6$, an overall factor which has a transition from constant to $\kappa^{-1}$ is found in Eq.~(\ref{III.B.6}).
Moreover, if $\alpha,\beta\gg30$, $\kappa$ is much larger than the width of the power spectrum peak until $\kappa\ll\kappa_-$ and $\ogw(k)$ enters the IR region. In this case, an extra transition from $\kappa^3\ln^2\kappa$ in the far IR region to $\kappa^2\ln^2\kappa$ in the near IR region occurs when $\kappa\sim(15/4)(\alpha\beta)^{-1/2}$.
This transition is also found in the previous works for a cut-off power spectrum and a log-normal power spectrum.

As for the IR and UV regions of $\ogw$, our results are compatible with the previous works.
The results for the IR case are given in Eqs.~(\ref{ogwIRA}), (\ref{ogwIRB}), (\ref{ogwIRC}) and (\ref{ogwIRD}) for different cases and are listed in Table \ref{tab:IR}. We see the typical $\kappa^3\ln^2\kappa$ and $\kappa^2\ln^2\kappa$ feature for $\alpha>3/2$ which was found in SIGWs induced by a general peaked power spectrum, and a $\kappa^{2\alpha}$ feature for $\alpha<3/2$ which is similar to the result of a power-law power spectrum. We also obtain a $\kappa^3\ln^3\kappa$ feature for the critical case $\alpha=3/2$. 
It is worth mentioning that the method in this part can be extended to more general cases, and the unique log-dependent scaling in the infrared region could be a smoking gun in searching for SIGWs. 
The behavior in UV region is given in Eqs.~(\ref{ogwUV}). The results on most areas are proportional to $\kappa^{-2\beta}$ while the $x^2-y^2\ll1$ region has a $\kappa^{-4-\beta}$ feature. For the critical case $\beta=4$, we also obtain a different $\kappa^{-8}\ln\kappa$ feature.

In this work, we have assumed that the scalar fluctuations are Gaussian, while non-Gaussianities are also important since the influence on PBH abundance and the SIGW spectrum can be significantly different. Moreover, we have only considered the tensor perturbations to the second order of scalar perturbations, while higher-order perturbations might lead to detectable signatures if the amplitude of the scalar power spectrum is pretty large.
The topics above may cause detectable deformation but the analytical results for more general cases still remain to be solved.

{\it Acknowledgments. } 
This work is supported by the National Key Research and Development Program of China Grant No.2020YFC2201502, grants from NSFC (grant No. 11991052, 12250010), Key Research Program of Frontier Sciences, CAS, Grant NO. ZDBS-LY-7009.
C.Y. acknowledge the financial support provided under the European Union’s H2020 ERC Advanced Grant “Black holes: gravitational engines of discovery” grant agreement no. Gravitas–101052587. Views and opinions expressed are however those of the author only and do not necessarily reflect those of the European Union or the European Research Council. Neither the European Union nor the granting authority can be held responsible for them. We acknowledge support from the Villum Investigator program supported by the VILLUM Foundation (grant no. VIL37766) and the DNRF Chair program (grant no. DNRF162) by the Danish National Research Foundation.

\appendix
\section{Features of broken power-law} \label{A.BPL}
In this section, we introduce some typical features of a broken power-law peak defined in Eq.~(\ref{PPL}):
\e
\mathcal{P_R}(\kappa) = A \frac{\alpha+\beta} {\beta\kappa^{-\alpha}+\alpha\kappa^\beta}.
\q
We will reduce it to a simpler form, and extend the approach to different models.

The most distinctive feature of a broken power-law spectrum is the power-law expansion on $\kappa=0$ and $\kappa\rightarrow+\infty$:
\begin{gather}
\mathcal{P_R}(\kappa) = A 
\begin{cases}
(1+\lambda) \kappa^{\alpha} \sum^\infty_{n=0}
(-1)^n K^n(\kappa) 
&\text{for}~K(\kappa)<1,\\
(1+\lambda^{-1}) \kappa^{-\beta} \sum^\infty_{n=0}
(-1)^n K^{-n}(\kappa)
&\text{for}~K(\kappa)>1,
\end{cases}
\label{III.A.1}\\
\delta_{IR} = (-1)^n K^{1+n}(\kappa) ,\quad
\delta_{UV} = (-1)^n K^{-1-n}(\kappa).
\end{gather}
$\lambda=\alpha/\beta$ and $K(\kappa)= \kappa^{\alpha+\beta}\lambda$ are introduced for simplicity, and $\delta$ is the relative error of the power expansion up to $K^n$ term. 
We assume that the power-law feature is broken when the leading order of the asymptotic expansion reaches the peak value, where $\kappa$ and $\mathcal{P_R}(\kappa)$ satisfy that
\begin{gather}
\kappa_-=\left( 1+ \lambda \right)^{-1/\alpha},~
\kappa_0=\lambda ^{-1/(\alpha+\beta)},~
\kappa_+=\left( 1+ \lambda^{-1} \right) ^{1/\beta},~
\kappa_- < \kappa_0 < \kappa_+ ,\notag\\
\mathcal{P_R}(\kappa_-) > \frac{A}{2},~
\mathcal{P_R}(\kappa_+) > \frac{A}{2},~
\mathcal{P_R}(\kappa_+)\mathcal{P_R}(\kappa_-) > \frac{A^2}{2}.\notag    
\end{gather}
Here $\kappa_-$ and $\kappa_+$ are the critical values of IR and UV expansion respectively, and $\kappa_0$ is the boundary of the convergence domain of Eq.~(\ref{III.A.1}). 
The relation indicates that even if we use the power-law tail directly and replace the transition part $\kappa_-<\kappa<\kappa_+$ of the power spectrum by peak value, the relative error is smaller than 1. 
Moreover, the relative error approaches 1 on one side only if the power index is much larger than that on the other side, i.e., $|\ln\lambda|\gg1$, and decays much more rapidly than the other side. 

In order to elevate the approximation to a higher accuracy, we introduce a variable transformation that maintains $\mathcal{P_R}$:
\e
\mathcal{P_R}(\alpha,\beta;\kappa) = 
\mathcal{P_R}(\lambda,1;\kappa^\beta) =
\mathcal{P_R}(\lambda^{-1},1;\kappa^{-\alpha}).
\q
Therefore, once the approximation of $\mathcal{P_R} (\lambda,1,\kappa^\beta)$ on $0<\kappa<1$ and $\lambda>0$ is finished, it can be extended to the complete parameter space $\alpha,\beta,\kappa>0$ with the accuracy preserved.
The selection is based on that $\kappa_-^\beta=(1+\lambda)^{-1/\lambda}>e^{-1}$, which make it convenient to extend the power-law tail to the peak with a quadratic polynomial. 
On the region $\kappa<\kappa_-$, the deviation of the power-law tail from $\mathcal{P_R}$ is roughly to the order $(2\lambda+1)$, and consequently a general $\mathcal{P_R}(\kappa)$ can be approximated by:
\m
\mathcal{P_R}(\lambda,1,\kappa) &\approx& A
\begin{cases}
(1+\lambda) \kappa^{\lambda} + 
(\mathcal{P_R}(\kappa_-)-1) \left(\kappa/\kappa_-\right) ^{(2\lambda+1) }
,& \text{for}~ \kappa < \kappa_-;\\
1 + (\mathcal{P_R}(\kappa_-)-1) \left[(\kappa-1) /(\kappa_- -1) \right]^2
,& \text{for}~ \kappa_-<\kappa< 1.
\end{cases}\label{PRlam}\\
\mathcal{P_R}(\alpha,\beta,\kappa) &=& A
\begin{cases}
\mathcal{P_R}(\lambda,1,\kappa^\beta), & \text{for}~ \kappa<1;\\
\mathcal{P_R}(\lambda^{-1},1,\kappa^{-\alpha}), & \text{for}~ \kappa>1.
\end{cases}\label{PRalp}
\n
Additionally, $\mathcal{P_R}^2(\kappa)$ will be used in the calculation of $\ogw$, since it is associated with the $p=|\boldsymbol{k}-\boldsymbol{p}|$ slice in Eq.~(\ref{II.3}). We use the similar approximation:
\m&&
\mathcal{P_R}^2(\alpha,\beta,\kappa) \approx A^2
\begin{cases}
(1+\lambda)^2 \kappa^{2\alpha}
\left( 1+ d_1 \kappa^{\beta+\alpha}
\right)^2,
& \text{for}~ \kappa < \kappa_-;\\
1 + d_2 \left( \kappa^\beta-1 \right)^2,
& \text{for}~ \kappa_- < \kappa < 1;\\
1 + d_3 \left( \kappa^{-\alpha}-1\right)^2
,& \text{for}~  1 < \kappa < \kappa_+; \\
(1+\lambda^{-1})^2 \kappa^{-2\beta}
\left(1 + d_4 \kappa^ {-\alpha-\beta}
\right)^2
,& \text{for}~ \kappa > \kappa_+.
\end{cases}\\&&
d_1=
\frac{\mathcal{P_R}(\kappa_-)-1} {\kappa_-^{\alpha+\beta} (1+\lambda)},~
d_2=
\frac{\mathcal{P_R}^2(\kappa_-)-1}{(\kappa_-^\beta-1)^2},~
d_3=
\frac{\mathcal{P_R}^2(\kappa_+)-1} {(\kappa_+^{-\alpha}-1)^2} ,~
d_4=
\frac{\mathcal{P_R}(\kappa_+)-1}{(1+\lambda^{-1})}
\kappa_+^ {\alpha+\beta}
\label{PR2coeff}.\n
We have used that $\kappa_-(\lambda,1) = \kappa_-^\beta(\alpha,\beta)$. Another integral will also be used:
\begin{align}
    \int_0^{+\infty}\mathcal{P_R}(t)\mathcal{P_R}(t^{-1})\mathrm{d}\ln t =
    \int_0^{+\infty}\frac{\alpha+\beta}{(\alpha+\beta s)(\alpha s+\beta)} \mathrm{d}s =
    \frac{2}{\sqrt{\alpha\beta}}\frac{\ln \lambda}{\lambda^{1/2}-\lambda^{-1/2}}.
\end{align}
The second factor of the result is not sensitive to $\lambda$, and the value in the limit of $\lambda\to1$ is simply 1.

Moreover, as have been mentioned, a peaked spectrum can always be regarded as broken power-law near the peak, and therefore Eqs. (\ref{PRlam})-(\ref{PR2coeff}) can be directly used for a near-peak approximation. The correspondence can be realized by match the power expansion at $\kappa=1$, and we enumerate several representative peak for example:
\begin{gather}
\mathcal{P}_{\mathrm{LN}}(k) =
A \exp{-\frac{\ln^2(k/k_\ast)} {2\Delta^2}} \approx
\mathcal{P}_{\mathrm{PL}} (\Delta^{-1},\Delta^{-1},k),\label{LN}\\
\mathcal{P}_{\mathrm{CLN}}(k) =
A \exp{-\lambda_1 \left(\frac{k}{k_\ast}-1-\ln \frac{k}{k_\ast}\right)- \lambda_2 \ln^2\frac{k}{k_\ast}} \approx 
\mathcal{P}_{\mathrm{PL}} \left( \frac{c_2-c_1}{2},\frac{c_2+c_1}{2},k \right)
,\end{gather}
where $c_1=\lambda_1/(\lambda_1+2\lambda_2),~ c_2=(8\lambda_2+4\lambda_1+4c_1^2)^{1/2}$, and furthermore, some specific limit gives more trivial spectrum:
\begin{gather}
\lim_{\alpha,\beta\rightarrow0} \mathcal{P_R} (\alpha, \beta ,\kappa) = A,\\
\lim_{\alpha\rightarrow\infty} \mathcal{P_R} ( \alpha , \frac{\alpha}{\kappa_-^{-\alpha}-1} , \kappa) = A ~\Theta (\kappa-\kappa_-)
,\\
\lim_{\alpha,\beta\rightarrow\infty}
\frac{\sqrt{\alpha\beta}}{\pi} 
\frac{\lambda^ {\frac{1}{2} \frac{\lambda-1}{\lambda+1}}}
{\csc\left(\pi/(\lambda+1)\right)}
\mathcal{P_R} ( \alpha, \beta , \kappa) = A ~ \delta(\kappa-1)
. \end{gather}
It can be verified that the approximation in Eqs.~(\ref{PRlam})-(\ref{PR2coeff}) maintain the limit, but it should be noticed that the confusion between the near-peak approximation and the limit to delta function will lead to a deviation considering the different decay modes.

\section{Contributions from region $x>\sqrt{6}/2$}\label{A.B}
We begin from Eqs.~(\ref{III.B.1}) and (\ref{III.B.2}).
If the enhancement in the power spectrum is relatively slow, we can expand the integrand in Eq.~(\ref{III.B.2}) up to $y^2/x^2$ order, and hence $I_y(x,q)$, where $q=\kappa x/\sqrt{2}$, can be derived analytically:
\begin{gather}
I_{y}(x,q) \approx \frac{8\sqrt{2}}{15} \left( 1+\frac{1}{14} \frac{4-\gamma(q)}{x^2} \right),\quad
\gamma(q)= \alpha\beta \frac{\beta +\left(\alpha ^2+2 \alpha  \beta +\alpha +(\beta -1) \beta \right) q^{\alpha +\beta }-\alpha  q^{2 (\alpha +\beta )}}{\left(\beta +\alpha  q^{\alpha +\beta }\right)^2}.
\label{III.B.4}\end{gather}
The above approximation is valid if $\alpha\beta\leq6$, which means that the width of the power spectrum $(\alpha\beta \kappa^2)^{-1/2}$ exceeds that of $(1-2y^2)^2$ before the peak leaves this region.
$\gamma(q)$ denotes a constant-order transition and we will use the asymptotic behavior that $\gamma(0)=\alpha$, $\gamma(1)=\alpha\beta$ and $\gamma(+\infty)=-\beta$, which are related to the IR tail, peak and the UV tail of the power spectrum respectively. 

According to the segmentation in Eq.~(\ref{PR2}), we use Eq.~(\ref{III.B.4}) directly with $\gamma(q)=\gamma(1)$ for the near-peak case $\kappa_-\leq q\leq\kappa_+$, and denote as $I_{y,p}(x)$:
\e I_{y,p}(x)=\frac{8\sqrt{2}}{15} \left( 1+\frac{1}{14} \frac{4-\alpha\beta}{x^2} \right). \label{Iyp1}\q
Then we connect the behavior of $\gamma(q)$ at $q=\sqrt{3}\kappa/2$ and $q\to+\infty$ to the near-peak approximation above. The selection is based on the boundary of the integration on $x$. In this case, $I_{y}(x,q)$ can be approximated to:
\begin{gather}
I_y(x,q) \approx
\begin{cases}
a_0 + a_{-2} x^{-2} 
& \text{for}~ x<\sqrt{2}\kappa_-/\kappa,\\
b_0 + b_{-2} x^{-2} 
& \text{for}~ \sqrt{2}\kappa_-/\kappa \leq x \leq \sqrt{2}\kappa_+/\kappa ,\\
c_0 + c_{-2} x^{-2} +c_{-4} x^{-4}
& \text{for}~ x > \sqrt{2}\kappa_+/\kappa,
\end{cases}~~ \label{Qxk1}
\\
a_{-2} =
\frac{I_{y,p}(\sqrt{2}\kappa_-/\kappa) - I_y(\sqrt{6}/2,\sqrt{3}\kappa/2) }{(\sqrt{2}\kappa_-/\kappa)^{-2}-2/3} ,~
a_{0}=I_{y,p}\(\sqrt{2}\kappa_-/\kappa\) - a_{-2} \(\frac{\sqrt{2}\kappa_-}{\kappa}\)^{-2} ,\notag\\
b_0=c_0=\frac{8\sqrt{2}}{15},~
b_{-2}=b_0\frac{1}{14}(4-\alpha\beta),~
c_{-2}=c_0\frac{1}{14}(4+\beta),~
c_{-4}=-\frac{c_0}{14}\beta(1+\alpha)
\left( \frac{\sqrt{2}\kappa_+}{\kappa}\right)^{2}.
\notag       
\end{gather}
The intervals correspond to $q<\kappa_-$, $\kappa_-\leq q\leq\kappa_+$, $q>\kappa_+$ respectively. The coefficients $a_{\rho}$, $b_{\rho}$ and $c_{\rho}$, where $\rho$ denote the corresponding power index of $x$, are relevant to $\kappa$ and the shape of power spectrum.
In the region $x<\sqrt{2}\kappa_-/\kappa$, we connect the value of $I_y$ at $x=\sqrt{6}/2$ to the near-peak case in Eq.~(\ref{Iyp1}), and the asymptotic behavior on $\kappa\rightarrow0$ will turn into Eq.~(\ref{III.B.4}) naturally with $\gamma(q)\to\alpha$.
Here the value of $I_y(\sqrt{6}/2,\sqrt{3}\kappa/2)$ will be given in Eq.~(\ref{IyNormal}). 
In the region $x>\sqrt{2}\kappa_+/\kappa$, we use Eq.~(\ref{III.B.4}) and introduce an extra $x^{-4}$ term to guarantee the continuity at $x=\sqrt{2}\kappa_+/\kappa$.

For the other case $\alpha\beta>6$, we focus on the peak value $I_y(x,1)$ first:
\m
I_y(x,1) &\approx& 
\left(\left(\frac{2 x}{\sqrt{\alpha  \beta -4}}\right)^{-2}+\left(\frac{8 \sqrt{2}}{15}\right)^{-2}\right)^{-1/2}.
\label{III.B.5}\n
We have connected the leading order of $2x\ll\sqrt{\alpha\beta-4}$ and $2x\gg\sqrt{\alpha\beta-4}$, with the relative error $<0.1$ compared with the normal approximation in Eq.~(\ref{IyNormal}). We use the expansion of Eq.~(\ref{III.B.5}) at $x=\sqrt{2}/\kappa$ as a near-peak approximation for $\kappa_-<q<\kappa_+$:
\begin{gather}
I_{y,p}(x) =
\left(\frac{2}{\kappa}\sqrt{\frac{2}{(\alpha\beta-4)(\Delta_\kappa+1)}}\right)
\frac{\Delta_\kappa+\kappa x/\sqrt{2}} {\Delta_\kappa+1}
,\quad \Delta_\kappa = \frac{225}{16}\frac{1}{\kappa^2(\alpha\beta-4)} . ~\label{III.B.6}   
\end{gather}
The feature of Eq.~(\ref{III.B.6}) depends on the value of $\Delta_\kappa$.
The value of the second factor of $I_{y,p}$ falls between $\kappa_-$ and $\kappa_+$ and simply reduces to 1 when $x=\sqrt{2}/\kappa$ or $\Delta_\kappa\gg1$, 
while the overall coefficient gives a transition from $\kappa^{-1}$ to constant near $\Delta_\kappa=1$, and the constant limit returns to that in Eq.~(\ref{III.B.4}). 

Similarly, we connect the asymptotic behavior to the near-peak approximation and the result for $\alpha\beta>6$ can be written as:
\begin{gather}
I_y(x,q) \approx
\begin{cases}
a_0 + a_{-2} x^{-2} 
& \text{for} ~ x<\sqrt{2}\kappa_-/\kappa ,\\
b_0 + b_1 x
& \text{for} ~ \sqrt{2}\kappa_-/\kappa \leq x \leq \sqrt{2}\kappa_+/\kappa,\\
c_0 + c_{-2} x^{-2} + c_{-4} x^{-4}
& \text{for} ~ x > \sqrt{2}\kappa_+/\kappa,
\end{cases}\label{Qxk2}\\
b_0=\frac{2}{\kappa}\sqrt{\frac{2}{(\alpha\beta-4)(\Delta_\kappa+1)}}\frac{\Delta_\kappa} {\Delta_\kappa+1},\quad
b_1=2\sqrt{\frac{1}{(\alpha\beta-4)(\Delta_\kappa+1)}} \frac{1} {\Delta_\kappa+1}
\notag\\
c_{-4} = \left(I_{y,p}(\sqrt{2}\kappa_+/\kappa)- c_0 - c_{-2} \left( \frac{\sqrt{2}\kappa_+}{\kappa}\right)^{-2} 
\right)
\left(\frac{\sqrt{2}\kappa_+}{\kappa}\right)^{4}.
\notag    
\end{gather}
$a_0$, $a_{-2}$, $c_0$ and $c_{-2}$ have the same form with that in Eq.~(\ref{Qxk1}).

For both $\alpha\beta\leq6$ and $\alpha\beta>6$, $I_y(x,q)$ has been reduced to the sum of a group of $x^{\rho}$ terms and the coefficients have been expressed as $a_{\rho}$, $b_{\rho}$ and $c_{\rho}$ which is related to IR, near-peak and UV region of the power spectrum respectively. 
Taking into account the approximation in Eq.~(\ref{PR2}), the contribution to $\ogw$ from this region can be written down directly by combining Eqs.~(\ref{PR2}) (\ref{Fxy}) (\ref{III.B.1}) (\ref{Qxk1}) and (\ref{Qxk2}):
\begin{subequations}\m
\ogw^{x>\frac{\sqrt{6}}{2}}(k) &=&
\ogw^{q<\kappa_-}(k)+\ogw^{\kappa_-<q<1}(k)+
\ogw^{1<q<\kappa_+}(k)+\ogw^{q>\kappa_+}(k),\label{Ogw1}\\
\ogw^{q<\kappa_-}(k) &\approx&  A^2(1+\lambda)^2
\sum_{\rho} \sum_{j=0}^2 a_{\rho} C_2^j d_1^j
\left(\frac{\kappa}{\sqrt{2}}\right)^ {2\alpha+j(\alpha+\beta)}
\Omega(\rho+2\alpha+j(\alpha+\beta),x)\Big{|}_{\sqrt{6}/2}^{R_1}, \label{Ogw1.1}\\
\ogw^{\kappa_-<q<1}(k) &\approx& A^2
\sum_{\rho} b_{\rho}\left(
\Omega(\rho,x)+d_2 \sum_{j=0}^2 C_2^j (-1)^j 
\left(\frac{\kappa}{\sqrt{2}}\right)^{j\beta}
\Omega(\rho+j\beta,x)\right) \Bigg{|}_{R_1}^{R_2},\label{Ogw1.2}\\
\ogw^{1<q<\kappa_+}(k) &\approx& A^2
\sum_{\rho} b_{\rho}\left(
\Omega(\rho,x)+d_3 \sum_{j=0}^2 C_2^j (-1)^j 
\left(\frac{\kappa}{\sqrt{2}}\right)^{-j\alpha}
\Omega(\rho-j \alpha,x)
\right) \Bigg{|}_{R_2}^{R_3},\label{Ogw1.3}\\
\ogw^{q>\kappa_+}(k) &\approx&  A^2(1+\lambda^{-1})^2
\sum_{\rho} \sum_{j=0}^2 c_{\rho} C_2^j d_4^j
\left(\frac{\kappa}{\sqrt{2}}\right)^ {-2\beta-j(\alpha+\beta)}
\Omega(\rho-2\beta-j(\alpha+\beta),x)
\Big{|}_{R_3}^{+\infty}\label{Ogw1.4}
.\n\end{subequations}

All the defined coefficients are listed below for convenience. Here, $C_m^n$ represents the binomial coefficients, $\lambda=\alpha/\beta$, and $R_1,~R_2$ and $R_3$ are defined as
\e
R_1=\max\{\frac{\sqrt{6}}{2},\frac{\sqrt{2}\kappa_-}{\kappa}\},~
R_2=\max\{\frac{\sqrt{6}}{2},\frac{\sqrt{2}}{\kappa}\},~ 
R_3=\max\{\frac{\sqrt{6}}{2},\frac{\sqrt{2}\kappa_+}{\kappa}\}.
\q
$\Omega(\rho,x)$ denotes the integration of $F(x,0)x^\rho$. Actually, the calculation can be done directly and the analytical results are listed in Appendix \ref{A.Orx}.

For the coefficients relevant to the segment of the scalar power spectrum which is derived in Appendix \ref{A.BPL}:
\begin{gather}
\mathcal{P_R}(q)=A \frac{\alpha+\beta} {\beta q^{-\alpha} + \alpha q^\beta},~
\kappa_-=\left( 1+ \lambda \right) ^{-1/\alpha},~
\kappa_+=\left( 1+ \lambda^{-1} \right) ^{1/\beta},\\
d_1=
\frac{\mathcal{P_R}(\kappa_-)-1} {\kappa_-^{\alpha+\beta} (1+\lambda)},~
d_2=
\frac{\mathcal{P_R}^2(\kappa_-)-1}{(\kappa_-^\beta-1)^2},~
d_3=
\frac{\mathcal{P_R}^2(\kappa_+)-1} {(\kappa_+^{-\alpha}-1)^2} ,~
d_4=
\frac{\mathcal{P_R}(\kappa_+)-1}{(1+\lambda^{-1})}
\kappa_+^ {\alpha+\beta}
.\end{gather}

Then for the coefficients relevant to the integration on $y$ direction, i.e. $I_y(x,q)$, the coefficients in each region is defined in Eq.~(\ref{Qxk1}) and (\ref{Qxk2}).
For the contribution from the region $\kappa_-\leq q \leq\kappa_+$, the relevant coefficients in Eq.~(\ref{Ogw1.2}) and Eq.~(\ref{Ogw1.3}) are that
\begin{gather}
(\rho,b_{\rho})\in
\begin{cases}
\left\{
\left(0,\dfrac{8\sqrt{2}}{15}\right),
\left(-2,\dfrac{b_0}{14}(4-\alpha\beta)\right)
\right\}
, &\text{for}~\alpha\beta\leq6;\\
\\
\left\{
\left(0, \sqrt{\dfrac{8}{(\alpha\beta-4)}} \dfrac{\kappa^{-1} \Delta_\kappa}{(\Delta_\kappa+1)^{3/2}} \right) ,
\left(1, \sqrt{\dfrac{4}{(\alpha\beta-4)}} \dfrac{1} {(\Delta_\kappa+1)^{3/2}} \right)
\right\}
, &\text{for}~\alpha\beta>6.
\end{cases}\label{brho}
\end{gather}
$\Delta_\kappa$ is defined as
\e
\Delta_\kappa = \frac{225}{16}\frac{1}{\kappa^2(\alpha\beta-4)}.
\q
For the region $q >\kappa_+$, the relevant coefficients in Eq.~(\ref{Ogw1.4}) are that
\m
(\rho,c_{\rho})\in\left\{
\left(0,\frac{8\sqrt{2}}{15} \right),
\left(-2,\frac{c_0}{14}(4+\beta) \right),
\left(-4, \left[ I_{y,p}\(\frac{\sqrt{2}\kappa_+}{\kappa}\) - c_0 - c_{-2} \left( \frac{\sqrt{2}\kappa_+}{\kappa}\right)^{-2} \right]
\left(\frac{\sqrt{2}\kappa_+}{\kappa}\right)^{4}\right)
\right\},\label{crho}
\n
where $I_{y,p}(x)$ is the near-peak approximation, which is expressed as
\m
I_{y,p}\(\frac{\sqrt{2}\kappa_\pm}{\kappa}\)=
\begin{cases}
\dfrac{8\sqrt{2}}{15} \left( 1+\dfrac{4-\alpha\beta}{14} \dfrac{\kappa^2}{2\kappa_\pm^2}
\right) &\text{for}~ \alpha\beta\leq6,\\ \\
\dfrac{2}{\kappa}\sqrt{\dfrac{2}{(\alpha\beta-4)(\Delta_\kappa+1)}}\dfrac{\Delta_\kappa+\kappa_\pm} {\Delta_\kappa+1} &\text{for}~ \alpha\beta>6,
\end{cases}\n
It is also used that for $\alpha\beta\leq6$, $c_{-4}$ reduces to that
\m
c_{-4}=-\frac{c_0}{14}\beta(1+\alpha)
\left( \frac{\sqrt{2}\kappa_+}{\kappa}\right)^{2}.
\n
For the region $q <\kappa_-$, the relevant coefficients in Eq.~(\ref{Ogw1.1}) are that
\m
(\rho,a_{\rho})\in\left\{
\left(0,
I_{y,p}\(\frac{\sqrt{2}\kappa_-}{\kappa}\) - \(\frac{\sqrt{2}\kappa_-}{\kappa}\)^{-2} a_{-2}\right),
\left(-2,
\frac{I_{y,p}(\sqrt{2}\kappa_-/\kappa) - I_y(\sqrt{6}/2,\sqrt{3}\kappa/2) }{(\sqrt{2}\kappa_-/\kappa)^{-2}-2/3} \right)
\right\}.\label{arho}
\n
The value of $I_y(\sqrt{6}/2,\sqrt{3}\kappa/2)$ is quite important here. Beginning from Eq.~(\ref{III.B.2}), we treat the peak of power spectrum on $y$ direction as normal and the result is that
\begin{gather}
I_y(x,q) \approx
\frac{\sqrt{\pi } \left(c^2+2 c+3\right) \text{erfi}\left(\sqrt{c/2}\right)}{c^{5/2}}-\frac{\sqrt{2} (c+3) e^{c/2}}{c^2}
\label{IyNormal},\\
c(x,q) = \frac{4-\gamma(q)}{x^2},~
\gamma(q)=
\alpha\beta
\frac{\beta +\left(\alpha ^2+2 \alpha  \beta +\alpha +(\beta -1) \beta \right) q^{\alpha +\beta }-\alpha  q^{2 (\alpha +\beta )}}{\left(\beta +\alpha  q^{\alpha +\beta }\right)^2},
\end{gather}
where $\text{erfi}(z)=-i \text{erf}(i z)$ denotes the imaginary error function.

\section{Analytical expressions of $\Omega(\rho,x)$}\label{A.Orx}
In this section, we give the analytical expression of $\Omega(\rho,x)$, which has been used in the calculation on $\ogw(k)$ in Eqs.~(\ref{Ogw1})-(\ref{Ogw1.4}), Appendix~\ref{A.B}.
$\Omega(\rho,x)$ is defined as
\begin{align}
\Omega(\rho,x)=&
\int\mathrm{d}x~ x^\rho F(x,0)\\=&
\frac{3\pi^2}{4}  \sum_{m=0}^4 \sum_{n=0}^2
C_4^m C_2^n (-3)^{4-m} (-2)^n I(-16+2m+2n+\rho,0,x) \notag\\&
+\frac{1}{9}\left(\frac{3}{2}\right)^{\frac{\rho-1}{2}}
\sum_{m=0}^2 \sum_{l=0}^2 \sum_{n=0}^{2+l} C_2^m C_{2+l}^n C_2^l (-3)^m (-2)^{4-n}
I(-\frac{13}{2}+\frac{\rho}{2}+m+n-l,l,\frac{2}{3}x^2),\label{ogwrho}    
\end{align}
where $I(a,l,z)$ is defined as
\m
I(a,l,z) &=& \int \mathrm{d}z\, z^a \ln^l{(z-1)}. \label{Ialz}
\n
The analytical expressions of the integral $I(a,l,z)$ is given below, where the case containing $m\in \mathbb{N}$ and $m>1$ is to cover the singularity at $a=-m$. 
\begin{subequations}
\m
I(a,2,z)&=&2(z-1)\,_4F_3(1,1,1,-a;2,2,2;1-z)-2(z-1) \ln (z-1) \,_3F_2(1,1,-a;2,2;1-z)\notag\\&&+ \frac{\left(z^{a+1}-1\right)\ln^2(z-1)}{a+1}, \label{alz1.1}\\
I(-1,2,z)&=&
-2 \text{Li}_3(1-z)+2 \text{Li}_2(1-z) \ln (z-1)+\ln (z) \ln ^2(z-1),
\\
I(a,1,z)&=&\frac{z^{a+2}  \Re \left[\,_2F_1(1,a+2;a+3;z)\right]} {(a+1)(a+2)}+\frac{z^{a+1}\ln(z-1)}{a+1},\\
I(-m,1,z)&=&\frac{1}{m-1}(\ln (z-1)-\ln (z)+\sum _{l=1}^{m-2} \frac{z^{-(-l+m-1)}}{-l+m-1}-z^{1-m} \ln (z-1)),\notag\\
&=& \frac{z^{1-m}}{1-m} \left(\Phi_L \left(z^{-1},1,m-1\right)+\ln (z-1)\right),\\
I(-1,1,z)&=&\text{Li}_2(1-z)+\ln (z-1) \ln (z),\\
I(a,0,z)&=&\frac{z^{a+1}}{a+1},~
I(-1,0,z)=\ln{z},\label{alz1.6}
\n
\end{subequations}
where $_p F_q$ is the hypergeometric function, $\text{Li}_n(x)$ is the polylog function and $\Phi_L$ is Lerch's transcendental function. The limits of the functions above are not trivial, so we give the results directly:  
\begin{subequations}
\m
\underset{z\rightarrow1+}{\lim}I(a,2,z)&=&0
,\label{alzlim1.1}\\
\underset{z\rightarrow+\infty}{\lim}I(a,2,z)&=&
-\frac{6 \left(H_{-a -2}\right)^2+6 \psi ^{(1)}(-a -1)+\pi ^2}{6 a +6}
,\\
\underset{z\rightarrow1+}{\lim}I(a,1,z)&=& -\frac{H_{a+1}}{a+1}, 
\\
\underset{z\rightarrow+\infty}{\lim}I(a,1,z)&=&
\frac{\pi \cot (\pi a)}{a+1}, 
\\
\underset{z\rightarrow1+}{\lim}I(-m,1,z)&=&
\frac{H_{m-2}}{m-1},
\\
\underset{z\rightarrow+\infty}{\lim}I(-m,1,z)&=&0, 
\\
\underset{z\rightarrow1+}{\lim}I(-1,1,z)&=&
0.\label{alzlim1.6}
\n
\end{subequations}
$H_n$ is the harmonic number and $\psi^{(1)}$ is the derivative of polygamma function.
It is also useful to exclude Eqs.~(\ref{alzlim1.1})-(\ref{alzlim1.6}) from Eqs.~(\ref{alz1.1})-(\ref{alz1.6}) when considering the asymptotic behavior, since the $I(a,l,z\gg1)$ will have a log-dependent $z^{a+1}$ feature automatically and the result will be the same as that in Appendix \ref{A.IR}.

\section{Details in IR region}\label{A.IR}
In Sec.~\ref{III.B}, we have given the dominant term of $\ogw(k)$ in IR region, i.e. $\kappa\ll\kappa_-$.
In this part, we give the detailed calculation and the more accurate result.
Further calculation is on the region $\kappa<\kappa_-\times10^{-1}$, where the approximation in Eq.~(\ref{FxyIR}) is valid in the second integral in Eq.~(\ref{ogwIR}).

Firstly, we give the asymptotic behavior of $\Omega(\rho,x)$ at $x\gg1$, which is used in our result in Sec. \ref{III.B}. The leading order of Eq.~(\ref{Ialz}) is that
\begin{align}
I(a,l,z\gg1)\approx\int \mathrm{d}z\, z^a\ln^l{z}=
\begin{cases}
\dfrac{z^{a+1}}{a+1}\ln^l{z}\(\sum_{j=0}^l \dfrac{\Gamma(l+1)}{\Gamma(l+1-j)}  
(-1)^{j} (1+a)^{-j}\ln^{-j}z\),
& \text{for}~a\neq-1;\\
\dfrac{1}{l+1}\ln^{l+1}{z}, & \text{for}~a=-1.
\end{cases}\label{Ialzinf}    
\end{align}
Note that the constants relevant to the asymptotic behavior in Eq.~(\ref{alz1.1})-(\ref{alz1.6}) have been subtracted from $I(a,l,z)$ for simplicity here.
The overall factor in the case $a\neq-1$ is the leading order of $\ln z$. Both circumstances have a log-dependent $z^{a+1}$ feature while the leading order of $\ln z$ is different. Using the approximation above, the leading order of Eq.~(\ref{ogwrho}) is that:
\m
\Omega(\rho,x\gg1)&\approx&
3\pi^2 I(-4+\rho,0,x) +
\left(\frac{3}{2}\right)^{\frac{\rho-1}{2}}
\sum_{l=0}^2  C_2^l (-2)^{2-l}
I(-\frac{5}{2}+\frac{\rho}{2},l,\frac{2}{3}x^2)
,\notag\\&=&
\begin{cases}
12\dfrac{x^{\rho-3}}{\rho-3}\left[\ln^2\left(\sqrt{\dfrac{2}{3}}e^ {\frac{\rho-2}{3-\rho }} x\right)+ 
\dfrac{\pi^2}{4}+\dfrac{1}{(\rho -3)^2}
\right] 
&\text{for}~\rho\neq3,\\
3\pi^2 \ln x +4\ln^3 \left(\sqrt{\dfrac{2}{3}} e^{-1}x \right)
&\text{for}~\rho=3.
\end{cases}
\label{ogwrhoinf}
\n
The result roughly shows a $x^{\rho-3}$ feature, but the dependence of $\ln x$ is different for $\rho=3$. 
Based on the approximation above, there are two properties which are helpful in further calculation of asymptotic behavior. The first is that
\m &&
\left(\frac{\kappa}{\sqrt{2}}\right)^\rho \Omega(\rho,x) \Big{|}_{\frac{\sqrt{2}}{\kappa}\kappa_i} ^{\frac{\sqrt{2}}{\kappa}\kappa_j} \approx
\begin{cases}
3\sqrt{2}\kappa^3 \left[ (\ln\kappa+B_1)^2+B_2 \right]\dfrac{\kappa_j^{\rho-3} -\kappa_i^{\rho-3}}{\rho-3}
,&\text{for}~\rho\neq3;\\\\
3\sqrt{2}\kappa^3 \left[ (\ln\kappa+B_1)^2+B_2 \right]\ln\dfrac{\kappa_j}{\kappa_i}
,&\text{for}~\rho=3.
\end{cases}
\label{k3ln2k}
\\&&
B_1=\begin{cases}
\ln\dfrac{\sqrt{3}}{2\sqrt{\kappa_i\kappa_j}} + \dfrac{\rho-2}{\rho-3}- \dfrac{1}{2} \dfrac{\kappa_j^{\rho-3}+\kappa_i^{\rho-3}}{\kappa_j^{\rho-3}-\kappa_i^{\rho-3}}\ln\dfrac{\kappa_j}{\kappa_i}
&\text{for}~\rho\neq3,\\
\ln\dfrac{\sqrt{3}}{2\sqrt{\kappa_i\kappa_j}}+1
&\text{for}~\rho=3;
\end{cases}\\&&
B_2=
\begin{cases}
\dfrac{\pi^2}{4}+\dfrac{1}{(\rho-3)^{2}}- \dfrac{\kappa_j^{\rho-3}\kappa_i^{\rho-3}}{\(\kappa_j^{\rho-3}-\kappa_i^{\rho-3}\)^2}\ln^2 \dfrac{\kappa_j}{\kappa_i}
, &\text{for}~\rho\neq3,\\
\dfrac{\pi^2}{4}+\dfrac{1}{12}\ln^2\dfrac{\kappa_j}{\kappa_i} &\text{for}~\rho=3.
\end{cases}
\n
In both cases the factor is proportional to $\kappa^3 \left[ (\ln\kappa+B_1)^2+B_2 \right]$ , as is mentioned in Eq.~(\ref{k3ln2k1}). 

Then if $\rho-3\gg1$ and $|\rho/\rho'|\gg1$ is satisfied, we have
\begin{align}
    \Omega(\rho,x)-x^{-\rho'}\Omega(\rho+\rho',x)\approx
    12x^{\rho-3}\frac{\rho'}{\rho^2}\[\frac{\pi^2}{4}    +\ln^2\(\sqrt{\frac{2}{3}}\frac{x}{e}\)\]\approx\frac{\rho'}{\rho}\Omega(\rho,x).
\end{align}

Based on the results above, we calculate the IR behavior of $\ogw^{x>\frac{\sqrt{6}}{2}}(k)$ part in Eq.~(\ref{ogw}) here. The detailed information has been listed in Eqs.~(\ref{Ogw1})-(\ref{Ogw1.4}), Appendix \ref{A.B}. It can be noticed that all of the terms has a similar $\kappa$-dependent factor 
\begin{align}
    \sum_\rho (a_\rho,b_\rho,c _\rho)\left(\frac{\kappa}{\sqrt{2}}\right)^{\rho'}\Omega(\rho+\rho',x),\label{Orxcoeff}
\end{align}
so that we will compare the factor for different part of $\ogw^{x>\frac{\sqrt{6}}{2}}(k)$ and find out the leading order in IR region.

\subsection{$\kappa_-<q<\kappa_+$ part}

Firstly we give the calculation on $\ogw^{\kappa_-<q<1}$ and $\ogw^{1<q<\kappa_+}(k)$ in Eqs.~(\ref{Ogw1.2}) and (\ref{Ogw1.3}). If $\alpha\beta\leq 6$, the coefficients $b_\rho$ are constant of $\kappa$, so that the factor in Eq.~(\ref{Orxcoeff}) is that
\begin{align}
    \sum_{\rho=0,-2} b_\rho\left(\frac{\kappa}{\sqrt{2}}\right)^{\rho'}\Omega(\rho+\rho',x)&= \frac{8\sqrt{2}}{15}
    \[\left(\frac{\kappa}{\sqrt{2}}\right)^{\rho'}   \Omega(\rho',x)+\frac{(4-\alpha\beta)\kappa^2}{28}\left(\frac{\kappa}{\sqrt{2}}\right)^{\rho'-2}\Omega(\rho'-2,x)\],\notag\\
    &\approx \frac{8\sqrt{2}}{15}
    \left(\frac{\kappa}{\sqrt{2}}\right)^{\rho'}   \Omega(\rho',x)\label{bOrx1}.
\end{align}
Using the approximation above, we can reduce Eqs.~(\ref{Ogw1.2}) and (\ref{Ogw1.3}) to that
\begin{align}
\ogw^{\kappa_-<q<1}(k) 
    \approx& \frac{8\sqrt{2}}{15}A^2 \left(\[\Omega(0,x)\]+d_2 \sum_{j=0}^2 C_2^j (-1)^j \[\left(\frac{\kappa}{\sqrt{2}}\right)^{j\beta}\Omega(j\beta,x)\] \right)\Bigg{|} _{\frac{\sqrt{2}\kappa_-}{\kappa}} ^{\frac{\sqrt{2}}{\kappa}},\notag\\
    \approx &\frac{16}{5}A^2[\kappa^3\ln^2\kappa] \left(\frac{\kappa_-^{-3}-1}{3}+d_2 \sum_{j=0}^2 C_2^j (-1)^j \frac{1-\kappa_-^{j\beta-3}}{j\beta-3} \right), \label{OgwIR1.2}\\
\ogw^{1<q<\kappa_+}(k)
    \approx&  \frac{8\sqrt{2}}{15}A^2 \left(\[\Omega(0,x)\]+d_3 \sum_{j=0}^2 C_2^j (-1)^j \[\left(\frac{\kappa}{\sqrt{2}}\right)^{-j\alpha} \Omega(-j \alpha,x)\] \right) \Bigg{|} _{\frac{\sqrt{2}}{\kappa}} ^{\frac{\sqrt{2}\kappa_+}{\kappa}},\notag\\
    \approx&\frac{16}{5}A^2[\kappa^3\ln^2\kappa] \left(\frac{1-\kappa_+^{-3}}{3}+d_3 \sum_{j=0}^2 C_2^j (-1)^j \frac{\kappa_+^{-j\alpha-3}-1}{-j\alpha-3} \right) .\label{OgwIR1.3}\\ 
\end{align}
The first step of the result is the linear combination of Eq.~(\ref{k3ln2k}), which is the complete result of $\kappa^3$ order.
In the last step, only the dominant terms $\kappa^3\ln^2\kappa$ are shown, which give more direct results.
In both cases, we put the $\kappa$-dependent factors into square brackets to make the result easier to understand.

As for $\alpha\beta>6$, the IR behavior of the coefficients $b_\rho$ is relevant to the value of $\Delta_{\kappa,\mathrm{IR}}$:
\begin{align}
    b_0\approx
    \begin{cases}
    \dfrac{8\sqrt{2}}{15}
    &\text{for}~\Delta_{\kappa,\mathrm{IR}}\gg1, \\\\
    \dfrac{8\sqrt{2}}{15}\Delta_{\kappa,\mathrm{IR}}^{3/2 }\propto\kappa^{-3}
    &\text{for}~\Delta_{\kappa,\mathrm{IR}}\ll1;
    \end{cases}\quad
    b_1\approx
    \begin{cases}
    \dfrac{8\sqrt{2}}{15}\Delta_{\kappa,\mathrm{IR}}^{-1}\dfrac{\kappa}{\sqrt{2}}\propto\kappa^3
    &\text{for}~\Delta_{\kappa,\mathrm{IR}} \gg1,\\\\
    \dfrac{8\sqrt{2}}{15}\Delta_{\kappa,\mathrm{IR}}^{1/2}\dfrac{\kappa}{\sqrt{2}}\propto\kappa^0 
    &\text{for}~\Delta_{\kappa,\mathrm{IR}} \ll1.
    \end{cases}
\end{align}
In IR region, we have showed in Eq.~(\ref{FxyIR}) that $F(x,y)$ barely influence $I_y(x,q)$, so that compared with Eq.~(\ref{III.B.6}), $\Delta_{\kappa,\mathrm{IR}}$ is defined as
\e
\Delta_{\kappa,\mathrm{IR}}= \frac{225}{16}\frac{1}{\kappa^2\alpha\beta}.\notag
\q

If $\Delta_{\kappa,\mathrm{IR}}\gg1$, the factor in Eq.~(\ref{Orxcoeff}) is that
\begin{align}
    \sum_{\rho=0,1} b_\rho\left(\frac{\kappa}{\sqrt{2}}\right)^{\rho'}\Omega(\rho+\rho',x)&= \frac{8\sqrt{2}}{15}\(\left(\frac{\kappa}{\sqrt{2}}\right)^{\rho'}\Omega(\rho',x)+\Delta_{\kappa,\mathrm{IR}}^{-1}\left(\frac{\kappa}{\sqrt{2}}\right)^{\rho'+1}\Omega(\rho'+1,x) \) ,\\&\approx
    \frac{8\sqrt{2}}{15}\left(\frac{\kappa}{\sqrt{2}}\right)^{\rho'}\Omega(\rho',x).
\end{align}
The result is the same as Eq.~(\ref{bOrx1}), which means that in this case, Eqs.~(\ref{Ogw1.2}) and (\ref{Ogw1.3}) have the same feature as Eqs.~(\ref{OgwIR1.2}) and (\ref{OgwIR1.3}). Note that since $\Delta_{\kappa,\mathrm{IR}}\gg1$ also include the case $\alpha\beta\leq6$ in IR region, we will consider the two cases together in the remaining part.

Then if $\Delta_{\kappa,\mathrm{IR}}\ll1$, the factor in Eq.~(\ref{Orxcoeff}) gives a different behavior that
\begin{align}
    \sum_{\rho=0,1} b_\rho\left(\frac{\kappa}{\sqrt{2}}\right)^{\rho'}\Omega(\rho+\rho',x)&= \frac{8\sqrt{2}}{15}\(\Delta_{\kappa,\mathrm{IR}}^{3/2} \left(\frac{\kappa}{\sqrt{2}}\right)^{\rho'}\Omega(\rho',x)+\Delta_{\kappa,\mathrm{IR}}^{1/2}\left(\frac{\kappa}{\sqrt{2}}\right)^{\rho'+1}\Omega(\rho'+1,x) \) ,\\&\approx
    \frac{8\sqrt{2}}{15}\Delta_{\kappa,\mathrm{IR}}^{1/2}\left(\frac{\kappa}{\sqrt{2}}\right)^{\rho'+1}\Omega(\rho'+1,x).
\end{align}
So that if $\Delta_{\kappa,\mathrm{IR}}\ll1$, we can reduce Eqs.~(\ref{Ogw1.2}) and (\ref{Ogw1.3}) to that
\begin{align}
    \ogw^{\kappa_-<q<1}(k) \approx& A^2\frac{8\sqrt{2}}{15} \[\Delta_{\kappa,\mathrm{IR}}^{1/2}\] \left(\[\frac{\kappa}{\sqrt{2}}\Omega(1,x)\] +d_2 \sum_{j=0}^2 C_2^j (-1)^j \[\left(\frac{\kappa}{\sqrt{2}}\right)^{1+j\beta}\Omega(1+j\beta,x)\]\right) \Bigg{|} _{\frac{\sqrt{2}\kappa_-}{\kappa}} ^{\frac{\sqrt{2}}{\kappa}},\notag\\
    \approx& A^2 \frac{12}{\sqrt{\alpha\beta}} \[\kappa^2\ln^2\kappa\] \left(\frac{\kappa_-^{-2}-1}{2} +d_2 \sum_{j=0}^2 C_2^j (-1)^j \frac{1-\kappa_-^{j\beta-2}}{j\beta-2}\right) ,\label{OgwIR2.2}\\
    \ogw^{1<q<\kappa_+}(k) \approx& A^2 \frac{8\sqrt{2}}{15} \[\Delta_{\kappa,\mathrm{IR}}^{1/2}\] \left(\[\frac{\kappa}{\sqrt{2}}\Omega(1,x)\] +d_3 \sum_{j=0}^2 C_2^j (-1)^j \[\left(\frac{\kappa}{\sqrt{2}}\right)^{1-j\alpha}\Omega(1-j \alpha,x)\]\right) \Bigg{|} _{\frac{\sqrt{2}}{\kappa}} ^{\frac{\sqrt{2}\kappa_+}{\kappa}},\notag\\
    \approx&  A^2 \frac{12}{\sqrt{\alpha\beta}} \[\kappa^2\ln^2\kappa\] \left(\frac{1-\kappa_+^{-2}}{2}+d_3 \sum_{j=0}^2 C_2^j (-1)^j \frac{\kappa_+^{-j\alpha-2}-1}{-j\alpha-2} \right) ,\label{OgwIR2.3}
\end{align}
Compared with Eqs.~(\ref{OgwIR1.2}) and (\ref{OgwIR1.3}), the results above have an extra $\Delta_{\kappa,\mathrm{IR}}^{1/2}$ factor, which is proportional to $\kappa^{-1}$. As a result, $\ogw^{\kappa_-<q<1}$ and $\ogw^{1<q<\kappa_+}(k)$ behaves as $\kappa^2\ln^2\kappa$ in the dominant term but the coefficients are much smaller.

\subsection{$q>\kappa_+$ part}
Next, we focus on the term $\ogw^{q>\kappa_+}$ defined in Eq.~({\ref{Ogw1.4}}). 
Firstly for $\Delta_{\kappa,\mathrm{IR}}\gg1$, the dominant term of Eq.~(\ref{Orxcoeff}) is that
\begin{align}
    \sum_{\rho=0,-2,-4} c_\rho\left(\frac{\kappa}{\sqrt{2}}\right)^{\rho'}\Omega(\rho+\rho',x)\approx
    \frac{8\sqrt{2}}{15}\left(\frac{\kappa}{\sqrt{2}}\right)^{\rho'}\Omega(\rho',x).
\end{align}
Applying the approximation to Eq.~({\ref{Ogw1.4}}), we have
\begin{align}
    \ogw^{q>\kappa_+}(k)
    \approx& A^2  (1+\lambda^{-1})^2 \frac{8\sqrt{2}}{15} \sum_{j=0}^2 C_2^j d_4^j \[\left(\frac{\kappa}{\sqrt{2}}\right)^ {-2\beta-j(\alpha+\beta)} \Omega(-2\beta-j(\alpha+\beta),x)\] \Bigg{|}_{\frac{\sqrt{2}\kappa_+}{\kappa}}^{+\infty},\notag\\
    \approx&\frac{16}{5}A^2(1+\lambda^{-1})^2 [\kappa^3\ln^2\kappa]  \sum_{j=0}^2 C_2^j d_4^j \frac{\kappa_+^{-3-2\beta-j(\alpha+\beta)}}{3+2\beta+j(\alpha+\beta)}.\label{OgwIR1.4}
\end{align}

Then if $\Delta_{\kappa,\mathrm{IR}}\ll1$, the coefficients $c_\rho$ are constant except $c_{-4}$, and its asymptotic behavior is that
\begin{equation}
    c_{-4}\approx
    \dfrac{8\sqrt{2}}{15}\kappa_+^{4}\left[\Delta_{\kappa,\mathrm{IR}}^{1/2}\kappa_+ -1-\dfrac{4+\beta}{14}\left( \dfrac{\sqrt{2}\kappa_+}{\kappa}\right)^{-2} \right]\left(\dfrac{\kappa}{\sqrt{2}}\right)^{-4}.
\end{equation}
We have used the IR behavior of $I_{y,p}(\sqrt{2}\kappa_\pm/\kappa)$:
\begin{equation}
    I_{y,p}(\sqrt{2}\kappa_\pm/\kappa)\approx
    \begin{cases}
    \dfrac{8\sqrt{2}}{15},&\text{for}~\Delta_{\kappa,\mathrm{IR}} \gg1;\\\\
    \dfrac{8\sqrt{2}}{15}\Delta_{\kappa,\mathrm{IR}}^{1/2}\kappa_\pm ,&\text{for}~\Delta_{\kappa,\mathrm{IR}} \ll1 .
    \end{cases}\label{IypIR}
\end{equation}
The dominant term of Eq.~(\ref{Orxcoeff}) is
\begin{align}
    \sum_{\rho=0,-2,-4} c_\rho\left(\frac{\kappa}{\sqrt{2}}\right)^{\rho'}\Omega(\rho+\rho',x)\approx
    \dfrac{8\sqrt{2}}{15}\kappa_+^{5} \Delta_{\kappa,\mathrm{IR}}^{1/2} \left(\dfrac{\kappa}{\sqrt{2}}\right)^{\rho'-4} \Omega(\rho'-4,x).
\end{align}
In this case, the IR behavior of $\ogw^{q>\kappa_+}$ is that
\begin{align}
\ogw^{q>\kappa_+}(k) \approx& A^2
 (1+\lambda^{-1})^2\kappa_+^{5} \frac{2\sqrt{2}}{\sqrt{\alpha\beta}} \[\kappa^{-1}\] \sum_{j=0}^2 C_2^j d_4^j \[\left(\frac{\kappa}{\sqrt{2}}\right)^ {-4-2\beta-j(\alpha+\beta)} \Omega(-4-2\beta-j(\alpha+\beta),x)\]\Bigg{|} _{\frac{\sqrt{2}\kappa_+}{\kappa}} ^{+\infty},\notag\\
    \approx& A^2
    (1+\lambda^{-1})^2\kappa_+^{5} \frac{12}{\sqrt{\alpha\beta}} \[\kappa^2\ln^2\kappa\] \sum_{j=0}^2 C_2^j d_4^j \frac{\kappa_+^{-7-2\beta-j(\alpha+\beta)}}{7+2\beta+j(\alpha+\beta)}\Bigg{|} _{\frac{\sqrt{2}\kappa_+}{\kappa}} ^{+\infty}\label{OgwIR2.4}.
\end{align}

\subsection{$q<\kappa_-$ part}
Finally for the the IR behavior of $\ogw^{q<\kappa_-}(k)$ part in Eq.~(\ref{Ogw1.1}), the coefficients $a_\rho$ listed in Eq.~(\ref{arho}) in the IR region can be expressed as
\begin{align}
a_0\approx I_{y,p}\(\frac{\sqrt{2}\kappa_-}{\kappa}\) - a_{-2}\(\frac{\sqrt{2}\kappa_-}{\kappa}\)^{-2},\quad
a_{-2}\approx -\frac{3}{2}\( I_{y,p}(\sqrt{2}\kappa_-/\kappa) - I_y(\sqrt{6}/2,0) \).
\end{align}
The asymptotic behavior of $I_y(\sqrt{6}/2,0)$ is simply a nonzero constant, and also $\propto\alpha^{-1/2}$ if $\alpha\gg1$. The IR behavior of $I_{y,p}(\sqrt{2}\kappa_-/\kappa)$ is given in Eq.~(\ref{IypIR}).

Firstly for $\Delta_{\kappa,\mathrm{IR}}\gg1$, the dominant term of Eq.~(\ref{Orxcoeff}) is that
\begin{align}
    \sum_{\rho=0,-2} a_\rho\left(\frac{\kappa}{\sqrt{2}}\right)^{\rho'}\Omega(\rho+\rho',x)\approx
    \frac{8\sqrt{2}}{15}\left(\frac{\kappa}{\sqrt{2}}\right)^{\rho'}\Omega(\rho',x).
\end{align}
In this case, Eq.~(\ref{ogw1.1}) in the limit of $\kappa\ll1$ gives
\m
\ogw^{q<\kappa_-}(k) &\approx&  A^2 (1+\lambda)^2 \Bigg[ \frac{8\sqrt{2}}{15}\sum_{j=0}^2 C_2^j d_1^j \left(\frac{\kappa}{\sqrt{2}}\right)^ {2\alpha+j(\alpha+\beta)} \Omega(2\alpha+j(\alpha+\beta), \frac{\sqrt{2}\kappa_-}{\kappa})\notag\\&&
-\left(\frac{\kappa}{\sqrt{2}}\right)^ {2\alpha} \sum_{\rho=0,-2} a_{\rho}\Omega(\rho+2\alpha,\frac{\sqrt{6}}{2}) \Bigg].\label{OgwIR1.1}
\n
If $\alpha<3/2$, the dominant term is the second term in the square bracket, i.e.
\begin{equation}
    \ogw^{q<\kappa_-}(k) \approx  -A^2 (1+\lambda)^2 \[ \left(\frac{\kappa}{\sqrt{2}}\right)^ {2\alpha}\] \sum_{\rho=0,-2} a_{\rho}\Omega(\rho+2\alpha,\frac{\sqrt{6}}{2}).
\end{equation}
In another case $\alpha>3/2$, the dominant term is the first term in the square bracket, i.e.
\begin{align}
    \ogw^{q<\kappa_-}(k) \approx&  A^2 (1+\lambda)^2 \frac{8\sqrt{2}}{15} \sum_{j=0}^2 C_2^j d_1^j \left(\frac{\kappa}{\sqrt{2}}\right)^ {2\alpha+j(\alpha+\beta)} \Omega(2\alpha+j(\alpha+\beta), \frac{\sqrt{2}\kappa_-}{\kappa}).\\
    \approx& \frac{16}{5} A^2 (1+\lambda)^2  \[\kappa^3\] \sum_{j=0}^2 C_2^j d_1^j \frac{\kappa_-^{2\alpha+j(\alpha+\beta)-3}}{2\alpha+j(\alpha+\beta)-3}\notag\\& \times\left[\left(\ln\frac{\kappa}{\kappa_-}
    +\frac{2\alpha+j(\alpha+\beta)-2}{2\alpha+j(\alpha+\beta)-3} \right) ^2+\dfrac{\pi^2}{4}+ \dfrac{1}{(2\alpha+j(\alpha+\beta)-3)^2}\right], \notag\\
    \approx& \frac{16}{5} A^2 (1+\lambda)^2 \[\kappa^3 \ln^2\kappa\]\sum_{j=0}^2 C_2^j d_1^j \frac{\kappa_-^{2\alpha+j(\alpha+\beta)-3}}{2\alpha+j(\alpha+\beta)-3}.\label{OgwIR1.1.2}
\end{align}
Considering $\alpha=3/2$ as a critical case, we find the result of $\ogw^{q<\kappa_-}(k)$ is different from Eq.~(\ref{OgwIR1.1.2}):
\m
\ogw^{q<\kappa_-}(k) &\approx& \frac{16}{5} A^2 (1+\lambda)^2 [\kappa^3] \Bigg\{
\[-\frac{1}{3}\ln^3\frac{\sqrt{3}e\kappa}{2\kappa_-}-\frac{\pi^2}{4}\ln\frac{\kappa}{\sqrt{2}\kappa_-} \]+
\sum_{j=1}^2 C_2^j d_1^j \frac{\kappa_-^{2\alpha+j(\alpha+\beta)-3}}{2\alpha+j(\alpha+\beta)-3}\notag\\&& \times\left[\left(\ln\frac{\kappa}{\kappa_-}
+\frac{2\alpha+j(\alpha+\beta)-2}{2\alpha+j(\alpha+\beta)-3} \right) ^2+\dfrac{\pi^2}{4}+ \dfrac{1}{(2\alpha+j(\alpha+\beta)-3)^2}\right]\Bigg\},\notag\\
&\approx&
\frac{16}{15} A^2 (1+\lambda)^2 \[-\kappa^3 \ln^3\kappa\].
\n
The dominant term turns into $\kappa^{3}\ln^3\kappa$.


Then if $\Delta_{\kappa,\mathrm{IR}}\ll1$, the the dominant term of Eq.~(\ref{Orxcoeff}) is that
\begin{align}
    \sum_{\rho=0,-2} a_\rho\left(\frac{\kappa}{\sqrt{2}}\right)^{\rho'}\Omega(\rho+\rho',x)\approx&
    \dfrac{8\sqrt{2}}{15}\Delta_{\kappa,\mathrm{IR}}^{1/2}\kappa_-\left(\frac{\kappa}{\sqrt{2}}\right)^{\rho'}\Omega(\rho',x).
\end{align}
In thas case, the IR behavior of Eq.~(\ref{Ogw1.1}) is that
\begin{align}
    \ogw^{q<\kappa_-}(k) \approx&  A^2 (1+\lambda)^2 \frac{2\sqrt{2}\kappa_-}{\sqrt{\alpha\beta}}\[\kappa^{-1}\] \sum_{j=0}^2 C_2^j d_1^j \[ \left(\frac{\kappa}{\sqrt{2}}\right)^ {2\alpha+j(\alpha+\beta)} \Omega(2\alpha+j(\alpha+\beta), \frac{\sqrt{2}\kappa_-}{\kappa})\], \notag\\
 \approx&  
A^2 (1+\lambda)^2 \frac{12}{\sqrt{\alpha\beta}} \kappa_-^{-2} \[\kappa^{2}\]\[\(\ln{\frac{\kappa}{\kappa_-}}+1\)^2+ \frac{\pi^2}{4}\] \sum_{j=0}^2 C_2^j d_1^j \frac{\kappa_+^{2\alpha+j(\alpha+\beta)}}{2\alpha+j(\alpha+\beta)-3},\notag\\
    \approx&A^2 (1+\lambda)^2 \frac{12}{\sqrt{\alpha\beta}} \kappa_-^{-2} \[\kappa^{2}\ln^2\kappa\] \sum_{j=0}^2 C_2^j d_1^j \frac{\kappa_+^{2\alpha+j(\alpha+\beta)}}{2\alpha+j(\alpha+\beta)-3}
\end{align}
and the dominant term is $\propto\kappa^2\ln^2\kappa$. In the three cases above corresponding to $\alpha\geq3/2$, the full result is relevant to $\ln(\kappa/\kappa_-)$, so that the dominant term, or in other words the last step of each case, can be taken as the full result if $-\ln\kappa\gg-\ln\kappa_->0$, or $\kappa\ll\kappa_-<1$.

\subsection{$x<\frac{\sqrt{6}}{2}$ part}
In Sec.~\ref{III.B}, we have dropped the terms $\ogw^{\frac{\sqrt{6}}{2}_-}(k)+ \ogw^{v\rightarrow0}(k)+\ogw^{\frac{\sqrt{2}}{2}_+}(k)$ in Eq.~(\ref{ogw}) from the IR behavior since their contribution is a magnitude lower. Anyway, we can reconsider this part to improve the precision. 

This part deserves consideration only when $\alpha\lesssim1$, and accounts for about 3\% of the result if $\alpha=1$. In this case, this part simply reduce to 
\begin{align}
    \ogw^{x<\frac{\sqrt{6}}{2}}(k) &\approx \int_\frac{1}{\sqrt{2}}^{\frac{\sqrt{6}}{2}} \mathrm{d}x \int_{-\frac{1}{\sqrt{2}}}^{\frac{1}{\sqrt{2}}} \mathrm{d}y \left(1-2y^2\right)^2 F(x,y) \mathcal{P_R}^2\(\frac{\sqrt{3}\kappa}{2}\),
    \\&\approx 0.18\times(1+\lambda)^2 \(\frac{\sqrt{3}\kappa}{2}\)^{2\alpha}.
\end{align}
We simply take $x=\frac{\sqrt{6}}{2},~y=0$ in the power spectrum since $\alpha$ is small, and $F(x,y)$ is greatly enhanced near $x=\frac{\sqrt{6}}{2}$.
In the second step, we have used the leading order of Eq.~(\ref{III.A.1}) as an approximation.

\smallskip
	
\bibliography{ref}
	
\end{document}